\input amssym.def


\magnification=\magstephalf
\hsize=14.0 true cm
\vsize=19 true cm
\hoffset=1.0 true cm
\voffset=2.0 true cm

\abovedisplayskip=12pt plus 3pt minus 3pt
\belowdisplayskip=12pt plus 3pt minus 3pt
\parindent=1em


\font\sixrm=cmr6
\font\eightrm=cmr8
\font\ninerm=cmr9

\font\sixi=cmmi6
\font\eighti=cmmi8
\font\ninei=cmmi9

\font\sixsy=cmsy6
\font\eightsy=cmsy8
\font\ninesy=cmsy9

\font\sixbf=cmbx6
\font\eightbf=cmbx8
\font\ninebf=cmbx9

\font\eightit=cmti8
\font\nineit=cmti9

\font\eightsl=cmsl8
\font\ninesl=cmsl9

\font\sixss=cmss8 at 8 true pt
\font\sevenss=cmss9 at 9 true pt
\font\eightss=cmss8
\font\niness=cmss9
\font\tenss=cmss10

 at 12 true pt
\font\bigrm=cmr10 at 12 true pt
\font\bigbf=cmbx10 at 12 true pt

 at 16 true pt
 at 16 true pt
 at 16 true pt

\catcode`@=11
\newfam\ssfam

\def\tenpoint{\def\rm{\fam0\tenrm}%
    \textfont0=\tenrm \scriptfont0=\sevenrm \scriptscriptfont0=\fiverm
    \textfont1=\teni  \scriptfont1=\seveni  \scriptscriptfont1=\fivei
    \textfont2=\tensy \scriptfont2=\sevensy \scriptscriptfont2=\fivesy
    \textfont3=\tenex \scriptfont3=\tenex   \scriptscriptfont3=\tenex
    \textfont\itfam=\tenit                  \def\it{\fam\itfam\tenit}%
    \textfont\slfam=\tensl                  \def\sl{\fam\slfam\tensl}%
    \textfont\bffam=\tenbf \scriptfont\bffam=\sevenbf
    \scriptscriptfont\bffam=\fivebf
                                            \def\bf{\fam\bffam\tenbf}%
    \textfont\ssfam=\tenss \scriptfont\ssfam=\sevenss
    \scriptscriptfont\ssfam=\sevenss
                                            \def\ss{\fam\ssfam\tenss}%
    \normalbaselineskip=13pt
    \setbox\strutbox=\hbox{\vrule height8.5pt depth3.5pt width0pt}%
    \let\big=\tenbig
    \normalbaselines\rm}

\def\ninepoint{\def\rm{\fam0\ninerm}%
    \textfont0=\ninerm      \scriptfont0=\sixrm
                            \scriptscriptfont0=\fiverm
    \textfont1=\ninei       \scriptfont1=\sixi
                            \scriptscriptfont1=\fivei
    \textfont2=\ninesy      \scriptfont2=\sixsy
                            \scriptscriptfont2=\fivesy
    \textfont3=\tenex       \scriptfont3=\tenex
                            \scriptscriptfont3=\tenex
    \textfont\itfam=\nineit \def\it{\fam\itfam\nineit}%
    \textfont\slfam=\ninesl \def\sl{\fam\slfam\ninesl}%
    \textfont\bffam=\ninebf \scriptfont\bffam=\sixbf
                            \scriptscriptfont\bffam=\fivebf
                            \def\bf{\fam\bffam\ninebf}%
    \textfont\ssfam=\niness \scriptfont\ssfam=\sixss
                            \scriptscriptfont\ssfam=\sixss
                            \def\ss{\fam\ssfam\niness}%
    \normalbaselineskip=12pt
    \setbox\strutbox=\hbox{\vrule height8.0pt depth3.0pt width0pt}%
    \let\big=\ninebig
    \normalbaselines\rm}

\def\eightpoint{\def\rm{\fam0\eightrm}%
    \textfont0=\eightrm      \scriptfont0=\sixrm
                             \scriptscriptfont0=\fiverm
    \textfont1=\eighti       \scriptfont1=\sixi
                             \scriptscriptfont1=\fivei
    \textfont2=\eightsy      \scriptfont2=\sixsy
                             \scriptscriptfont2=\fivesy
    \textfont3=\tenex        \scriptfont3=\tenex
                             \scriptscriptfont3=\tenex
    \textfont\itfam=\eightit \def\it{\fam\itfam\eightit}%
    \textfont\slfam=\eightsl \def\sl{\fam\slfam\eightsl}%
    \textfont\bffam=\eightbf \scriptfont\bffam=\sixbf
                             \scriptscriptfont\bffam=\fivebf
                             \def\bf{\fam\bffam\eightbf}%
    \textfont\ssfam=\eightss \scriptfont\ssfam=\sixss
                             \scriptscriptfont\ssfam=\sixss
                             \def\ss{\fam\ssfam\eightss}%
    \normalbaselineskip=10pt
    \setbox\strutbox=\hbox{\vrule height7.0pt depth2.0pt width0pt}%
    \let\big=\eightbig
    \normalbaselines\rm}

\def\tenbig#1{{\hbox{$\left#1\vbox to8.5pt{}\right.\n@space$}}}
\def\ninebig#1{{\hbox{$\textfont0=\tenrm\textfont2=\tensy
                       \left#1\vbox to7.25pt{}\right.\n@space$}}}
\def\eightbig#1{{\hbox{$\textfont0=\ninerm\textfont2=\ninesy
                       \left#1\vbox to6.5pt{}\right.\n@space$}}}

\font\sectionfont=cmbx10
\font\subsectionfont=cmti10

\def\figurecaptionfont{\ninepoint}
\def\tablecaptionfont{\ninepoint}
\def\footnotefont{\eightpoint}


\newcount\equationno
\newcount\bibitemno
\newcount\figureno
\newcount\tableno

\equationno=0
\bibitemno=0
\figureno=0
\tableno=0
\advance\pageno by -1


\footline={\ifnum\pageno=0{\hfil}\else
{\hss\rm\the\pageno\hss}\fi}


\def\section #1. #2 \par
{\vskip0pt plus .20\vsize\penalty-100 \vskip0pt plus-.20\vsize
\vskip 1.6 true cm plus 0.2 true cm minus 0.2 true cm
\global\def\equationlabel{#1}
\global\equationno=0
\leftline{\sectionfont #1. #2}\par
\immediate\write\terminal{Section #1. #2}
\vskip 0.7 true cm plus 0.1 true cm minus 0.1 true cm
\noindent}


\def\subsection #1 \par
{\vskip0pt plus 0.8 true cm\penalty-50 \vskip0pt plus-0.8 true cm
\vskip2.5ex plus 0.1ex minus 0.1ex
\leftline{\subsectionfont #1}\par
\immediate\write\terminal{Subsection #1}
\vskip1.0ex plus 0.1ex minus 0.1ex
\noindent}


\def\appendix #1 \par
{\vskip0pt plus .20\vsize\penalty-100 \vskip0pt plus-.20\vsize
\vskip 1.6 true cm plus 0.2 true cm minus 0.2 true cm
\global\def\equationlabel{\hbox{\rm#1}}
\global\equationno=0
\leftline{\sectionfont Appendix #1}\par
\immediate\write\terminal{Appendix #1}
\vskip 0.7 true cm plus 0.1 true cm minus 0.1 true cm
\noindent}


\def\enum{\global\advance\equationno by 1
(\equationlabel.\the\equationno)}


\def\ifundefined#1{\expandafter\ifx\csname#1\endcsname\relax}

\def\ref#1{\ifundefined{#1}?\immediate\write\terminal{unknown reference
on page \the\pageno}\else\csname#1\endcsname\fi}

\newwrite\terminal
\newwrite\bibitemlist

\def\bibitem#1#2\par{\global\advance\bibitemno by 1
\immediate\write\bibitemlist{\string\def
\expandafter\string\csname#1\endcsname
{\the\bibitemno}}
\item{[\the\bibitemno]}#2\par}

\def\beginbibliography{
\vskip0pt plus .15\vsize\penalty-100 \vskip0pt plus-.15\vsize
\vskip 1.2 true cm plus 0.2 true cm minus 0.2 true cm
\leftline{\sectionfont References}\par
\immediate\write\terminal{References}
\immediate\openout\bibitemlist=biblist
\frenchspacing\parindent=1.5em
\vskip 0.5 true cm plus 0.1 true cm minus 0.1 true cm}

\def\endbibliography{
\immediate\closeout\bibitemlist
\nonfrenchspacing\parindent=1.0em}


\def\figurecaption#1{\global\advance\figureno by 1
\narrower\figurecaptionfont
Fig.~\the\figureno. #1}

\def\tablecaption#1{\global\advance\tableno by 1
\vbox to 0.5 true cm { }
\centerline{\tablecaptionfont%
Table~\the\tableno. #1}
\vskip-0.4 true cm}

\tenpoint

\immediate\openin\bibitemlist=biblist
\ifeof\bibitemlist\immediate\closein\bibitemlist
\else\immediate\closein\bibitemlist
\input biblist \fi


\def\thisyear{\number\year}

\def\thismonth{\ifcase\month\or
January\or February\or March\or April\or May\or June\or
July\or August\or September\or October\or November\or December\fi}



\def\rmd{{\rm d}}
\def\rmD{{\rm D}}
\def\rme{{\rm e}}
\def\rmO{{\rm O}}
\def\rmU{{\rm U}}


\def\rz{{\Bbb R}}
\def\gz{{\Bbb Z}}


\def\proof{\noindent{\sl Proof:}\kern0.6em}
\def\endproof{\hskip0.6em plus0.1em minus0.1em
\setbox0=\null\ht0=5.4pt\dp0=1pt\wd0=5.3pt
\vbox{\hrule height0.8pt
\hbox{\vrule width0.8pt\box0\vrule width0.8pt}
\hrule height0.8pt}}
\def\frac#1#2{\hbox{$#1\over#2$}}
\def\dual{\mathstrut^*\kern-0.1em}
\def\mod{\;\hbox{\rm mod}\;}
\def\ring{\mathaccent"7017}
\def\lvec#1{\setbox0=\hbox{$#1$}
    \setbox1=\hbox{$\scriptstyle\leftarrow$}
    #1\kern-\wd0\smash{
    \raise\ht0\hbox{$\raise1pt\hbox{$\scriptstyle\leftarrow$}$}}
    \kern-\wd1\kern\wd0}
\def\rvec#1{\setbox0=\hbox{$#1$}
    \setbox1=\hbox{$\scriptstyle\rightarrow$}
    #1\kern-\wd0\smash{
    \raise\ht0\hbox{$\raise1pt\hbox{$\scriptstyle\rightarrow$}$}}
    \kern-\wd1\kern\wd0}


\def\nab#1{{\nabla_{#1}}}
\def\nabstar#1{\nabla\kern-0.5pt\smash{\raise 4.5pt\hbox{$\ast$}}
               \kern-4.5pt_{#1}}
\def\drv#1{{\partial_{#1}}}
\def\drvstar#1{\partial\kern-0.5pt\smash{\raise 4.5pt\hbox{$\ast$}}
               \kern-5.0pt_{#1}}

\def\ldrvstar#1{\lvec{\,\partial}\kern-0.5pt\smash{\raise 4.5pt\hbox{$\ast$}}
               \kern-5.0pt_{#1}}




\def\gfields{{\frak U}}
\def\group{{\frak G_0}}
\def\cfields{{\frak U_0}}
\def\vfields{{\frak A}}


\def\psibar{\overline{\psi}}
\def\anomaly{{\cal A}}
\def\jinfty{j\kern0.5pt\smash{\raise 4.0pt\hbox{$\scriptstyle\star$}}
            \kern-5.0pt}
\def\Atrans{A^{\raise0pt\hbox{\sixrm T}}}
\def\atrans{A^{\raise0pt\hbox{\fiverm T}}}

\def\Etrans{\eta^{\raise0pt\hbox{\sixrm T}}}
\def\etrans{\eta^{\raise0pt\hbox{\fiverm T}}}
\def\Elong{\eta^{\raise0pt\hbox{\sixrm L}}}
\def\elong{\eta^{\raise0pt\hbox{\fiverm L}}}

\def\Ztrans{\zeta^{\raise0pt\hbox{\sixrm T}}}
\def\ztrans{\zeta^{\raise0pt\hbox{\fiverm T}}}
\def\Zlong{\zeta^{\raise0pt\hbox{\sixrm L}}}
\def\zlong{\zeta^{\raise0pt\hbox{\fiverm L}}}


\def\dirac#1{\gamma_{#1}}
\def\diracstar#1#2{
    \setbox0=\hbox{$\gamma$}\setbox1=\hbox{$\gamma_{#1}$}
    \gamma_{#1}\kern-\wd1\kern\wd0
    \smash{\raise4.5pt\hbox{$\scriptstyle#2$}}}
\def\hatdirac#1{\hat{\gamma}_{#1}}
\def\hatdiracstar#1#2{
    \setbox0=\hbox{$\gamma$}\setbox1=\hbox{$\gamma_{#1}$}
    \hat{\gamma}_{#1}\kern-\wd1\kern\wd0
    \smash{\raise4.5pt\hbox{$\scriptstyle#2$}}}


\def\tr{{\rm tr}}
\def\Tr{{\rm Tr}}


\def\L{{\frak L}}
\def\Linfty{\L\kern0.5pt\smash{\raise 4.5pt\hbox{$\scriptstyle\star$}}
            \kern-4.0pt}
\def\Wline{{\frak W}}
\def\Bfield{{\frak B}}
\def\bfield{{\frak b}}
\def\Gfield{{\frak C}}

\def\Lapprox{{\frak K}}
\def\S{{\frak S}}
\def\R{{\frak R}}


\def\trans{{\cal Q}}
\def\trfun{\tau}


\def\weight{w}


\def\chigamma{Q_{\Gamma}} 

\rightline{DESY 98-180}

\vskip 3.0 true cm minus 0.3 true cm
\centerline
{\bigbf Abelian chiral gauge theories on the lattice}
\vskip 1.5ex
\centerline
{\bigbf with exact gauge invariance}
\vskip 1.5 true cm
\centerline{\bigrm Martin L\"uscher}
\vskip1ex
\centerline{\it Deutsches Elektronen-Synchrotron DESY}
\centerline{\it Notkestrasse 85, D-22603 Hamburg, Germany}
\centerline{\it E-mail: luscher@mail.desy.de}
\vskip 2.5 true cm
\centerline{\bf Abstract}
\vskip 2.0ex
It is shown that U(1) chiral gauge theories with anomaly-free
multiplets of Weyl fermions can be put on the lattice
without breaking the gauge invariance or violating any other
fundamental principle.
The Ginsparg-Wilson relation plays a key r\^ole in this 
construction, which is non-perturbative and 
includes all topological sectors of the theory in finite volume.
In particular, 
the cancellation of the gauge anomaly and 
the absence of global topological obstructions
can be established on the basis of this relation and
the lattice symmetries alone.
\vfill
\centerline{\thismonth\space\thisyear}
\eject

\section 1. Introduction

For well-known reasons the 
formulation of chiral gauge theories on the lattice proves to be
difficult and no completely satisfactory
solution of the problem has been found so far [\ref{Shamir}].
One of the propositions that have been made is to put
the gauge-fixed theory on the lattice and to include a set of 
counterterms in the action with coefficients chosen in such a way
that the BRS symmetry is restored in the continuum limit~[\ref{Rome}]
(for a review and further references see 
refs.~[\ref{RomeReview},\ref{BuckowReview}]).
Using lattices with different lattice spacings for the 
gauge and the fermion fields
is another idea which is being actively pursued [\ref{IntA}--\ref{IntD}].
The symmetry breaking terms can then be suppressed
by choosing the lattice spacing in the fermion sector
to be much smaller than the other lattice spacing.

A few years ago
an entirely different approach was suggested 
by Kaplan [\ref{Kaplan}], who noted 
that fermion modes which are bound to a 
four-dimensional defect in a five-dimensional lattice
are chiral under certain conditions.
Later this led to the so-called overlap formulation of chiral gauge 
theories [\ref{Overlap},\ref{OverlapII}],
where the fermion partition function is written
as a transition matrix element (the ``overlap")
between the ground states of two auxiliary Hamilton operators.
This development no doubt represents 
a big step forward,
but as in the other cases the gauge symmetry is broken
on the lattice. Moreover the locality properties of the 
theory are not transparent.

In this paper we consider U(1) gauge theories
where the gauge field couples to $N$ left-handed Weyl fermions with 
charges $\rme_{\alpha}$ satisfying
$$
  \sum_{\alpha=1}^N\,\rme_{\alpha}^3=0.
  \eqno\enum
$$
This is the classical condition for anomaly cancellation 
and the continuum theory is thus well-defined
to all orders of perturbation theory.
In particular, the effective gauge field action generated by
the fermions is
uniquely determined up to finite renormalizations of the gauge coupling
[\ref{Leutwyler}--\ref{EriceLectures}].

The main result obtained here is that these theories can be put
on the lattice without breaking the gauge invariance or violating
other basic principles such as the requirement of locality. 
The construction is non-perturbative and 
one has the right number and type of Weyl fermions from the beginning.
Not surprisingly it is technically rather
involved and perhaps not as explicit as one would like,
particularly in finite volume, where the non-trivial topology 
of the space of gauge fields gives rise
to additional complications.
The present paper
is hence mainly of theoretical interest, clarifying a question of principle,
but it does not provide a formulation of chiral gauge theories on the 
lattice which would be immediately usable for non-perturbative
studies through numerical simulations.
One should however note that 
chiral gauge theories are anyway a difficult case
for numerical simulations,
because the effective action has a non-zero imaginary part in general.

The starting point in this paper is the recent discovery 
that chiral symmetry can be preserved on the lattice without having 
to compromise in any other ways [\ref{HasenfratzI}--\ref{LuscherI}].
One achieves this by choosing a lattice Dirac operator $D$
satisfying 
\footnote{$\dagger$}{\footnotefont
For notational convenience the lattice spacing $a$ is set to $1$ 
so that all length scales are given 
in numbers of lattice spacings. In particular, the right-hand side
of eq.~(1.2) should be multiplied with $a$ if physical units are employed}
$$
  \dirac{5}D+D\dirac{5}=D\dirac{5}D.
  \eqno\enum
$$
This relation
(which is originally due to Ginsparg and Wilson [\ref{GinspargWilson}])
guarantees that the fermion action is invariant
under a group of infinitesimal transformations which may be
regarded as a lattice form of the usual chiral symmetries.
Moreover the non-invariance of the fermion integration measure
under flavour-singlet transformations straightforwardly leads to the 
expected axial anomaly [\ref{LuscherI},\ref{KikukawaYamada}].

Having an exact chiral symmetry of the action, it turns out to be
relatively easy to introduce left- and right-handed fields
[\ref{Niedermayer}]. Because of the anomaly the
fermion integration measure however does not decompose in a unique way
and one ends up with a gauge field dependent phase ambiguity.
To fix the phase of the measure
so that the gauge symmetry and the locality of the theory
are preserved is the principal problem which has to solved 
if one would like to set up chiral gauge theories along 
these lines. 

All this will be explained in more detail in the next two sections.
We then discuss the conditions which an ideal
fermion integration measure should fulfil (sect.~4).
Whether such measures exist is far from obvious
and the rest of the paper is in fact entirely devoted to this question.
For clarity the results are first presented in sect.~5 in a concise
form, with all proofs and technical details being deferred
to sections 6--11.
A few concluding remarks are collected in sect.~12.

\section 2. Lattice fields and functional integral

The lattice theories constructed in this paper are defined
in the traditional manner, where one begins by 
specifying the space of fields and the lattice action.
Expectation values of arbitrary products of the fields are then
obtained as usual from the functional integral.
The definition of the integration measure for Weyl fermions 
is non-trivial, however, and there are further technical details
which need to be discussed carefully.
A summary of notational conventions is included in appendix~A.

\subsection 2.1 Gauge fields

We choose lattice units and construct the theory
on a finite lattice of size $L$ with periodic boundary conditions.
U(1) gauge fields on such a lattice may be
represented through periodic link fields,
$$
  \eqalignno{
  U(x,\mu)&\in {\rm U(1)},
  \qquad 
  x=(x_0,\ldots,x_3)\in\gz^4, 
  &\enum\cr
  \noalign{\vskip2ex}
  U(x+L\hat{\nu},\mu)&=U(x,\mu) 
  \quad\hbox{for all}\quad
  \mu,\nu=0,\ldots,3,
  &\enum\cr}
$$
on the infinite lattice. The independent degrees of freedom
are then the link variables at the points $x$ in the block
$$
  \Gamma=\left\{x\in\gz^4\bigm| 0\leq x_{\mu}<L\right\}
  \eqno\enum
$$
($L\geq1$ is assumed to be an integer).
Under gauge transformations
$$
  U(x,\mu)\to \Lambda(x)U(x,\mu)\Lambda(x+\hat{\mu})^{-1},
  \eqno\enum
$$
the periodicity of the field will be preserved 
if $\Lambda(x)\in{\rm U(1)}$ is periodic.
This is not the most general possibility, but the convention is here
adopted that only periodic functions are referred to as gauge transformations.

For the gauge field action $S_{\rm G}$ we take a somewhat unusual expression
which effectively imposes an upper bound on the lattice field tensor. 
The reasons for this will become clear later.
As in the case of the standard Wilson action we write
$$
  S_{\rm G}={1\over4g_0^2}\sum_{x\in\Gamma}\sum_{\mu,\nu}\,
  {\cal L}_{\mu\nu}(x)
  \eqno\enum
$$
with $g_0$ being the bare coupling. 
The plaquette action is however taken to be of the more complicated form
$$
  {\cal L}_{\mu\nu}(x)=\cases{
  \left[F_{\mu\nu}(x)\right]^2
  \bigl\{1-\left[F_{\mu\nu}(x)\right]^2/\epsilon^2\bigr\}^{-1}
  &if $|F_{\mu\nu}(x)|<\epsilon$,\cr
  \noalign{\vskip1.5ex}
  \infty & otherwise,\cr
  \noalign{\vskip0.6ex}}
  \eqno\enum
$$
where $\epsilon$ is a fixed number in the range $0<\epsilon<\frac{1}{3}\pi$
and the field tensor $F_{\mu\nu}(x)$ is defined through
$$
  \eqalignno{
  F_{\mu\nu}(x)&=
  {1\over i}\ln P(x,\mu,\nu),
  \qquad
  -\pi<F_{\mu\nu}(x)\leq\pi,
  &\enum\cr\noalign{\vskip1.5ex}
  P(x,\mu,\nu)&=U(x,\mu)U(x+\hat{\mu},\nu)
  U(x+\hat{\nu},\mu)^{-1}U(x,\nu)^{-1}.
  &\enum\cr}
$$
Note that the Boltzmann factor $\rme^{-S_{\rm G}}$ 
is a smooth function of the link 
variables with this choice of action.
In particular, the functional integral can be set up in 
the usual way with the standard integration measure for U(1) lattice
gauge fields.

This concludes the definition of the pure gauge part of the 
theory. There are a few remarks which should be added here.

\vskip1ex
\noindent
(a)~The Boltzmann factor is a product of local factors,
one for each plaquette on the lattice. The locality of the theory
is thus guaranteed. Moreover since it is differentiable,
no special precautions are required when performing
partial integrations in the functional integral 
(such as those needed when deriving the field equations). 

\vskip1ex
\noindent
(b)~As already mentioned, our choice of action is such that
the functional integral is effectively restricted to the space of fields 
satisfying
$$
  |F_{\mu\nu}(x)|<\epsilon
  \quad\hbox{for all $x,\mu,\nu$.}
  \eqno\enum
$$
Gauge fields of this type will be referred to as
{\it admissible}\/ in the following. 

\vskip1ex
\noindent
(c)~When physical units are employed, the parameter $\epsilon$
should be replaced by $\epsilon/a^2$ where $a$ denotes the lattice spacing.
It is then immediately clear that the curly bracket in the definition (2.6)
of the action and the bound (2.9)
are irrelevant in the classical continuum limit.
As far as the weak coupling phase is concerned,
there is in fact little doubt that
the lattice theory defined here is 
in the same universality class as the standard lattice theory.
In particular, it is a valid
lattice regularization of the free U(1) gauge theory.

\subsection 2.2 Magnetic flux sectors

Before proceeding to the fermion fields, 
we briefly discuss the topology of the space
of admissible fields.
Proofs and further details will be given in sect.~7.
One might expect that there is no interesting topological structure
in this simple theory, but this is not so. 
The key observation is that the magnetic flux
$$
  \phi_{\mu\nu}(x)=
  \sum_{s,t=0}^{L-1}\,F_{\mu\nu}(x+s\hat{\mu}+t\hat{\nu})
  \eqno\enum
$$
through the $(\mu,\nu)$--planes of the lattice is conserved and 
quantized. In other words, for any admissible field
the associated flux satisfies
$$
  \phi_{\mu\nu}(x)=2\pi m_{\mu\nu},
  \eqno\enum
$$
where $m_{\mu\nu}=-m_{\nu\mu}$ is an integer tensor independent of $x$.

Evidently the flux quantum numbers $m_{\mu\nu}$ cannot change
when the gauge field is continuously deformed. The field space
is thus a disjoint union of the sectors
of all admissible fields with a given magnetic flux. 
Moreover it can be shown that 
each of these sectors has the topology a multi-dimensional torus
times a convex space.

\subsection 2.3 Lattice Dirac operator

We first consider Dirac fermions and discuss
the projection to the left-handed components in the next subsection.
Dirac fields $\psi(x)$ on the lattice carry a Dirac index
and a flavour index $\alpha=1,\ldots,N$.
As in the case of the gauge field it is convenient to assume
that the fermion fields are defined on the infinite lattice.
Periodic boundary conditions are then imposed through
the requirement that
$$
  \psi(x+L\hat{\mu})=\psi(x)
  \quad\hbox{for all}\quad\mu=0,\ldots,3.
  \eqno\enum
$$
Other types of periodic boundary conditions could be
admitted here with little change in the following.

Under gauge transformations $\Lambda(x)$ the 
fermion fields transform according to the representation 
$$
  \psi(x)\to R\left[\Lambda(x)\right]\psi(x),
  \qquad
  R\left[\Lambda(x)\right]_{\alpha\beta}=
  \delta_{\alpha\beta}\,\Lambda(x)^{\rme_{\alpha}},
  \eqno\enum
$$
where $\rme_{\alpha}\in\gz$ is the charge of the fermion
with flavour $\alpha$. Throughout the paper we take it for granted
that the condition for anomaly cancellation, eq.~(1.1), is satisfied.
A simple example of an acceptable charge assignment is thus given by
$$
  \rme_1=\rme_2=\ldots=\rme_8=1,
  \qquad
  \rme_9=-2.
  \eqno\enum
$$
Taking pairs of charges with opposite sign is another possibility, 
but in the present context this is a less interesting case,
because one ends up with a chiral theory which is 
effectively vector-like.

The proper choice of the lattice Dirac operator $D$ is 
of central importance in the following. Apart from
being a solution of the Ginsparg-Wilson relation (1.2),
the operator should fulfil a number of technical requirements.
In particular, it should be local, gauge covariant and 
differentiable in the gauge field.
The complete list of requirements is given in appendix B.

Gauge covariant solutions of the 
Ginsparg-Wilson relation are not easy to find.
The ``perfect" lattice Dirac operator of 
refs.~[\ref{HasenfratzI},\ref{HasenfratzII}] is one of them
and another solution has been derived by Neuberger [\ref{NeubergerI}] from
the overlap formalism.
In this case all properties described in appendix B have
been established rigorously [\ref{Locality}].
Note that it suffices to
define the Dirac operator for all admissible gauge fields since only 
these contribute to the functional integral.
The relevant results of ref.~[\ref{Locality}] in fact apply
for admissible fields only and if 
$\epsilon$ is such that 
$|\rme_{\alpha}|\,\epsilon<\frac{1}{30}$ for all $\alpha$.

In infinite volume the action of the Dirac operator is given by 
$$
  D\psi(x)=\sum_{y\in\gz^4}D(x,y)\psi(y),
  \eqno\enum
$$
where the kernel $D(x,y)$ is a matrix in Dirac and flavour space.
For periodic fields eq.~(2.15) may be rewritten in the form
$$
  D\psi(x)=\sum_{y\in\Gamma}\,D_L(x,y)\psi(y),
  \qquad
  D_L(x,y)=\sum_{n\in\gz^4}D(x,y+Ln),
  \eqno\enum
$$
i.e.~the finite-volume kernel $D_L(x,y)$ is obtained from
the kernel on the infinite lattice by applying the reflection principle.
Evidently, since we are dealing with the same operator,
the Ginsparg-Wilson relation holds in finite volume too.

From the properties listed in appendix B it follows
that $D_L(x,y)$ is periodic in $x$ and $y$ separately.
Moreover it transforms in the same way as $D(x,y)$ 
under the gauge and lattice symmetries and from the 
locality of the operator one infers that
$$
  D_L(x,y)=D(x,y)+\rmO\bigl(\rme^{-L/\varrho}\bigr),
  \eqno\enum
$$ 
where $\varrho$ is the localization range of $D$.

\subsection 2.4 Weyl fermions

Using the Ginsparg-Wilson relation, the infinitesimal transformation 
$$
  \delta\psi=\dirac{5}(1-D)\psi,
  \qquad 
  \delta\psibar=\psibar\dirac{5},
  \eqno\enum
$$
is easily shown to be a symmetry of the fermion action  
$$
  S_{\rm F}=\sum_{x\in\Gamma}\,
  \psibar(x)D\psi(x).
  \eqno\enum
$$
Eq.~(2.18) is the chiral transformation of ref.~[\ref{LuscherI}]
except that the fermion 
and the anti-fermion fields are here treated asymmetrically. 
The reason for this is that the present formulation 
allows one to decompose the fields into left- and right-handed components
in a natural way.
First note that the operator $\hatdirac{5}=\dirac{5}(1-D)$ 
satisfies
$$
  (\hatdirac{5})^{\dagger}=\hatdirac{5},
  \qquad
  (\hatdirac{5})^{2}=1.
  \eqno\enum
$$
So if we define the projectors
$$ 
  \hat{P}_{\pm}=\frac{1}{2}(1\pm\hatdirac{5}),
  \qquad
  P_{\pm}=\frac{1}{2}(1\pm\dirac{5}),
  \eqno\enum
$$ 
it is immediately clear that the left-handed fields
$\hat{P}_{-}\psi$ and $\psibar P_{+}$ (and the complementary components)
transform under lattice chiral rotations in the same way as 
the corresponding fields in the continuum theory.
In particular, left- and right-handed fields decouple in the action (2.19)
\footnote{$\dagger$}
{\footnotefont 
That the action can be split this way
has independently been noted
by Hasenfratz and Niedermayer [\ref{HasenfratzNiedermayer},\ref{Niedermayer}].
A closely related observation has also been made by Narayanan
in the context of the 
overlap formalism [\ref{OverlapSplit}]}.

We now eliminate the right-handed components by 
imposing the constraints
$$
  \hat{P}_{-}\psi=\psi,
  \qquad
  \psibar P_{+}=\psibar,
  \eqno\enum
$$
on the fermion fields. An important point to note here is that these
conditions are local and gauge-invariant. The same is true
for the action (2.19) and we thus have a completely satisfactory
definition of the theory at the classical level.

\subsection 2.5 Fermion integration measure

To set up the quantum theory we also need to specify
an integration measure for left-handed fields.
The basic difficulty which one has here is 
that the subspace of left-handed fermion fields
depends on the gauge field. As a consequence there is a
non-trivial phase ambiguity in the integration measure.

To make this explicit let us suppose that $v_j(x)$, $j=1,2,3,\ldots\,$, is 
a basis of complex valued, periodic fermion fields such that
$$
  \hat{P}_{-}v_j=v_j,
  \qquad
  (v_k,v_j)=\delta_{kj}
  \eqno\enum
$$
(the bracket denotes the obvious scalar product for fermion fields in 
finite volume). 
The quantum field may then be expanded according to
$$
  \psi(x)=\sum_j\,v_j(x)c_j,
  \eqno\enum
$$
where the coefficients $c_j$ generate a Grassmann algebra.
They represent the independent degrees of freedom of the field
and an integration measure for left-handed fermion
fields is thus given by
$$
  \rmD[\kern0.5pt\psi\kern0.5pt]=\prod_{j}\,\rmd c_j.
  \eqno\enum
$$
An important mathematical fact which should be kept in mind 
in the following is that the measure is independent of the particular
basis that has been chosen up to a phase factor.
One can quickly see this by noting that a change of basis
$$
  \tilde{v}_j(x)=\sum_l\,v_l(x)(\trans^{-1})_{lj},
  \qquad
  \tilde{c}_j=\sum_l\,\trans_{jl}c_l,
  \eqno\enum
$$
implies a change of the measure by the factor $\det\trans$
which is a pure phase factor since $\trans$ is unitary. 
On the other hand, the remark shows that one has a phase ambiguity
which is cannot be ignored because the basis (and hence the phase 
of the measure) depends on the gauge field.
One can try to fix the ambiguity in some ad hoc manner,
but as will become clear in sect.~4
such prescriptions are likely to be unsatisfactory.
For the time being we assume that some particular basis has been chosen
and proceed with the definition of the theory.

In the case of the anti-fermion fields 
the subspace of left-handed fields
is independent of the gauge field and one can take the same
orthonormal basis $\bar{v}_k(x)$ for all gauge fields.
The ambiguity in the integration measure
$$
  \rmD[\kern0.5pt\psibar\kern0.5pt]=\prod_k\,\rmd \bar{c}_k,
  \qquad
  \psibar(x)=\sum_k\,\bar{c}_k\bar{v}_k(x),
  \eqno\enum
$$
is then only a constant phase factor.

Fermion expectation values of any product $\cal O$ of 
fields are now obtained as usual through the functional integral
$$
  \langle{\cal O}\rangle_{\rm F}=\weight[m]
  \int\rmD[\kern0.5pt\psi\kern0.5pt]\rmD[\kern0.5pt\psibar\kern0.5pt]
  \,{\cal O}\,\rme^{-S_{\rm F}}.
  \eqno\enum
$$
Note that this integral is completely well-defined.
The integration variables are the coefficients 
$c_j$ and $\bar{c}_k$ in terms of which the action 
assumes the form
$$
  S_{\rm F}=\sum_{k,j}\,\bar{c}_k M_{kj}c_j,
  \qquad
  M_{kj}=\sum_{x\in\Gamma}\,\bar{v}_k(x)Dv_j(x).
  \eqno\enum
$$
The fermion fields in the product $\cal O$
should be expanded similarly and 
the integral can then be evaluated following the standard rules
for Grassmann integration. 

In the definition (2.28) a complex
factor $\weight[m]$ has been included, which allows one to 
adjust the relative phase and absolute weight of the 
topological sectors. The factor only depends on the
magnetic flux quantum numbers $m_{\mu\nu}$ and we are free
to set $w[0]=1$. One might be tempted to do the same in all other sectors
as well, but this may not be the proper choice since the 
number of integration variables depends on the sector which is being
considered (cf.~subsect.~3.2). 

Full normalized expectation values 
are finally given by
$$
  \langle{\cal O}\rangle={1\over{\cal Z}}
  \int \rmD[U]\,\rme^{-S_{\rm G}}\langle{\cal O}\rangle_{\rm F},
  \eqno\enum
$$
where the normalization factor $\cal Z$ is defined through 
the requirement that $\langle1\rangle=1$ and $\rmD[U]$ 
denotes the usual integration measure for U(1) lattice gauge fields.

\section 3. Correlation functions and effective action

Apart from the fact that we have not fixed the phase of the fermion
integration measure, the theory is completely defined at this point
and one can begin to study its properties. In the following paragraphs
we work out a few quantities and address some of the basic questions
which one may have in order to demonstrate the consistency of the approach.

\subsection 3.1 Fermion propagator

If $D$ has no zero-modes it is straightforward to show that  
$$
  \langle\psi(x)\psibar(y)\rangle_{\rm F}=
  \langle 1\rangle_{\rm F}\times
  \hat{P}_{-}S_L(x,y)P_{+},
  \eqno\enum
$$
where the fermion propagator $S_L(x,y)$ is a periodic function
satisfying
$$
  \sum_{y\in\Gamma}\,D_L(x,y)S_L(y,z)=
  \delta_{xz}
  \quad\hbox{for all}\quad x,z\in\Gamma.
  \eqno\enum
$$
In other words, $S_L(x,y)$ is the kernel of the inverse of the Dirac
operator in finite volume. Note that there is no dependence
on the bases $v_j(x)$ and $\bar{v}_k(x)$ here since the phase
ambiguity of the fermion integration measure cancels in eq.~(3.1).

From the above and the definition of the chiral projectors
it follows that
$$
  \hat{P}_{-}S_L(x,y)P_{+}=P_{-}S_L(x,y)P_{+}+\frac{1}{2}P_{+}\delta_{xy}
  \eqno\enum
$$
for all points $x,y$ in $\Gamma$. This expression makes it evident
that the propagating fermion modes are chiral.
The theory thus describes the right number and type of Weyl fermions
and there is little doubt that one recovers
the correct Feynman rules in the continuum limit for the 
propagator in an external field.

\subsection 3.2 Fermion number violation

A characteristic feature of 
chiral gauge theories is that fermion number violating processes can take 
place. This is possible whenever the numbers of 
left- and right-handed zero-modes
of the Dirac operator, $n_{-}$ and $n_{+}$, are not the same.

We can now easily check this in the lattice theory.
First note that the dimensions of the 
spaces of left-handed fermion and anti-fermion fields
can be different. Since these spaces are the eigenspaces of
the corresponding chiral projectors, the
difference of their dimensions is given by
\footnote{$\dagger$}{\footnotefont
Here and below
the symbol ``$\Tr_L$" implies a trace over the space of fermion fields
in finite volume, ``$\Tr$" the same in infinite volume and ``$\tr$"
a trace over Dirac and flavour indices only}
$$
  \Tr_L\{\hat{P}_{-}\}-
  \Tr_L\{P_{+}\}=
  \frac{1}{2}\kern0.5pt\Tr_L\{\dirac{5}D\}=n_{-}-n_{+},
  \eqno\enum
$$
where the second equality follows from the index theorem
[\ref{HasenfratzII},\ref{LuscherI}]. 
The index $n_{+}-n_{-}$ is a topological invariant which assumes
a fixed and in general non-zero value in each magnetic flux sector.

In all sectors where the index does not vanish,
the matrix $M_{kj}$ which appears in the action (2.29) has
thus a rectangular shape. So if we temporarily choose
the basis vectors $v_j(x)$ and $\bar{v}_k(x)$ such that the 
first of them are the zero-modes, the action becomes
$$
  S_{\rm F}=\sum_{\kern4pt k>n_{+}}\sum_{\kern4pt j>n_{-}}\,
  \bar{c}_kM_{kj}c_j,
  \eqno\enum
$$
which is a non-degenerate quadratic form in the integration 
variables $c_j$ and $\bar{c}_k$ associated with the other modes.
The functional integral (2.28) hence vanishes unless
$\cal O$ is a product of $n_{-}$ fermion and $n_{+}$ anti-fermion
fields times an arbitrary polynomial in pairs of these fields 
and the gauge field variables. In other words, $\cal O$ has to have
a net fermion number equal to $n_{-}-n_{+}$ and the lattice theory
thus complies with the expected selection rules for 
fermion number violating processes.

\subsection 3.3 Effective action

In the vacuum sector
the dimensions of the spaces of 
left-handed fermion and anti-fermion fields are the same
and the fermion partition function is hence given by
$$
  \langle1\rangle_{\rm F}=\det M.
  \eqno\enum
$$
Chiral determinants in the continuum theory are usually 
studied by computing their variation under infinitesimal deformations 
of the gauge field [\ref{Leutwyler}--\ref{EriceLectures}]. 
We can do the same here and it will soon become clear that this 
is a useful exercise. 

So let us consider a variation 
$$
  \delta_{\eta}U(x,\mu)=i\eta_{\mu}(x)U(x,\mu)
  \eqno\enum
$$ 
of the gauge field, where $\eta_{\mu}(x)$ is any real periodic vector field.
After some algebra the associated variation of the effective action 
is then found to be given by
$$
  \delta_{\eta}\ln\det M=\Tr_L\{\delta_{\eta}D \hat{P}_{-}D^{-1}P_{+}\}
  -i\L_{\eta}.
  \eqno\enum
$$
One might have expected to end up with the first term only, 
but since the basis vectors $v_j(x)$ depend 
on the gauge field one has a second term,
$$ 
  \L_{\eta}=i\sum_j\,(v_j,\delta_{\eta}v_j),
  \eqno\enum
$$ 
which may be regarded as a contribution of 
the fermion integration measure.

The current $j_{\mu}(x)$ which is defined through
$$
  \L_{\eta}=\sum_{x\in\Gamma}\,\eta_{\mu}(x)j_{\mu}(x)
  \eqno\enum
$$
is going to play an important r\^ole in the following.
In particular, it will be shown later that the measure
can be reconstructed from the current if certain conditions are fulfilled.
Note that
the measure term transforms according to 
$$
  \widetilde{\L}_{\eta}=\L_{\eta}-i\delta_{\eta}\ln\det\trans
  \eqno\enum
$$
under basis transformations (2.26) and $\L_{\eta}$ is hence unchanged
if the transformation 
preserves the integration measure. As a consequence the current should 
be thought of as a quantity which is associated with the measure
rather than the basis vectors $v_j(x)$. It is also immediately clear
from this that any two measures with the same current 
are related to each other by a constant phase factor in each topological sector.

\subsection 3.4 Integrability condition

The significance of the measure term $\L_{\eta}$ may be further elucidated
by computing the ``curvature" 
$\delta_{\eta}\L_{\zeta}-\delta_{\zeta}\L_{\eta}$.
Starting from eq.~(3.9),
this is easily done and in a few lines one obtains
$$
  \delta_{\eta}\L_{\zeta}-\delta_{\zeta}\L_{\eta}
  =i\,\Tr_L\bigl\{\hat{P}_{-}
  \bigl[\delta_{\eta}\hat{P}_{-},\delta_{\zeta}\hat{P}_{-}\bigr]\bigr\}.
  \eqno\enum
$$
As expected from the transformation law (3.11), the curvature 
does not depend on the choice of the basis vectors $v_j(x)$. 
In particular, if it is not equal to zero
it cannot be made to vanish by adjusting the 
basis and in these cases the measure term is hence required
to ensure the integrability of eq.~(3.8).

It is interesting to note in this connection 
that essentially the same happens
in Leutwyler's construction of the chiral determinant in the 
continuum theory [\ref{Leutwyler}], where a local counterterm
has to be added to restore the integrability after applying 
a finite-part prescription to the variation of the determinant.
The analogy will be even more striking
after the discussion in the next section,
which will lead us to require that the current $j_{\mu}(x)$
should be a local expression in the gauge field.
The measure term then assumes the form of a local counterterm.

\subsection 3.5 Gauge anomaly

Although the fermion action and the projection to the left-handed
fields are gauge-invariant, the effective action tends to be 
non-invariant due to the anomaly and the fact that 
the fermion integration measure depends on the gauge field.
To work this out, let us consider a gauge variation
$$
  \eta_{\mu}(x)=-\drv{\mu}\omega(x),
  \eqno\enum
$$
where $\omega(x)$ is any periodic gauge function
and $\drv{\mu}$ the forward difference operator 
defined in appendix A. If we introduce the generator
$$
  T_{\alpha\alpha'}=\delta_{\alpha\alpha'}\kern0.5pt\rme_{\alpha}
  \eqno\enum
$$
of the fermion representation (2.13) of the gauge group,
it is then obvious that
$$
  \delta_{\eta}D=i\left[\omega T,D\right]
  \eqno\enum
$$ 
and taking eq.~(3.8) into account one obtains
$$
  \eqalignno{
  \delta_{\eta}\ln\det M&=i\sum_{x\in\Gamma}\,\omega(x)
  \left\{\anomaly_L(x)-\drvstar{\mu}j_{\mu}(x)\right\},
  &\enum\cr
  \noalign{\vskip1.5ex}
  \anomaly_L(x)&=-\frac{1}{2}\kern1.0pt\tr\left\{\dirac{5}TD_L(x,x)\right\},
  &\enum\cr}
$$ 
for the gauge variation of the effective action. Note that
$\anomaly_L(x)$ is equal to the sum of the axial anomalies 
associated with the $N$ flavours of fermions in the theory,
weighted with their charge
[\ref{HasenfratzII},\ref{LuscherI}].
In other words, $\anomaly_L(x)$ is the anomaly of the current
which couples to the gauge field.

\subsection 3.6 Vector-like theories

If the charges $\rme_{\alpha}$ come in pairs with opposite sign,
the continuum theory is formally equivalent to a vector-like theory,
where the gauge field couples to $\frac{1}{2}N$ Dirac fermions
with positive charges.
On the lattice we can choose a basis of left-handed fermion fields
such that the basis vectors in the sectors with positive and negative charges
are related to each other through
$$
  v_j^{-}(x)=\dirac{5}C^{-1}[v^{+}_j(x)]^{\ast},
  \eqno\enum
$$
where $C$ denotes the charge conjugation matrix.
The associated fermion integration measure 
is the same for any such basis and it is also easy to show that 
the measure term $\L_{\eta}$ vanishes.

If the basis $\bar{v}_k(x)$ of left-handed anti-fermion fields
is taken to be of the same type, 
the partition function (3.6) factorizes and 
in a few lines one obtains
$$
  \langle1\rangle_{\rm F}=
  \Bigl|\kern1.0pt\det_{\rme_{\alpha}>0} M\kern1.0pt\Bigr|^2=
  \det_{\rme_{\alpha}>0}D,
  \eqno\enum
$$
which is the expected result for a vector-like theory.
Up to contact terms and with an appropriate assignment of field components,
there is in fact a complete matching between
the chiral and the vector theory in the vacuum sector.
Presumably this is also the case in the other sectors,
but the issue will not be pursued here.

\section 4. Conditions on the fermion integration measure

According to the universality hypothesis,
the details of the lattice theory should not influence 
the continuum limit, apart from finite renormalizations, as long as 
a few basic principles are respected.
One of them is that the theory should be formulated locally
with no long-range couplings in the action. Symmetries are also
very important and the universality of the continuum limit
is more likely to be guaranteed if they are preserved on the lattice.

The conditions on the fermion integration measure listed below
have been devised with this in mind. 
They should be regarded as a maximal set of requirements
which one may reasonably hope to fulfil and 
one can be quite confident that the correct
continuum limit will be obtained if they are all satisfied.

\vskip2ex\noindent
(1)~{\it Differentiability with respect to the gauge field.}
The expectation value $\langle{\cal O}\rangle_{\rm F}$ 
of arbitrary (finite) products $\cal O$ of the fermion 
fields and the link variables should be smooth functions 
of the gauge field. 
This is a somewhat technical requirement,
but there are a few instances where 
the smoothness of the fermion integrals seems to be essential.
In particular, the derivation of the field equation discussed below is 
invalid if this is not guaranteed.

As explained in sect.~8, this condition assumes a simple 
form in terms of the basis vectors $v_j(x)$ and 
in the following we shall say that the fermion integration 
measure is smooth if the chosen basis has the properties stated there.

\vskip2ex\noindent
(2)~{\it Locality of the field equations.}
In euclidean field theory the field equations 
are linear relations between operator insertions in 
correlation functions. 
If the action is local, these operators are local composite fields
and the locality properties of the theory are thus directly
reflected by the field equations. 

This leads us to require that the fermion integration measure
should be such that the locality of the field equations is guaranteed.
In particular, this should be so for the 
field equations associated with the gauge field,
which one derives from the functional integral (2.30) by 
calculating the change of the integrand under field variations
of the type considered in subsect.~3.3. 
Relatively little work is required for this if the field product
$\cal O$ does not involve the fermion fields, because
only the sectors with vanishing index 
contribute in this case and one can then
make use of eqs.~(3.8) and (3.1) to show that
$$
  \langle\{\delta_{\eta}S_{\rm G}+
  \raise1pt\hbox{$\scriptstyle\sum_{x\in\Gamma}$}
  \psibar(x)\delta_{\eta}D\psi(x)+i\L_{\eta}\}{\cal O}\rangle
  =\langle\delta_{\eta}{\cal O}\rangle.
  \eqno\enum
$$
For local variations $\eta_{\mu}(x)$ the first two terms 
in this equation are manifestly local. 
To ensure the locality of the field equations we thus require that 
the current $j_{\mu}(x)$ is a local function of the gauge field
\footnote{$\dagger$}{\footnotefont
The notion of locality used in this paper is the same as in 
refs.~[\ref{Locality},\ref{Niedermayer},\ref{LuscherII}].
Details are given in appendix B for the case of the Dirac operator.
Note that the term only makes sense if 
the lattice size $L$ is much larger than
the localization range of the fields that one is interested in
}.

If one considers more general field products $\cal O$, the field 
equations are not quite as easy to derive, but in all cases it turns out
that the locality of the current implies the locality
of the field equations.

\vskip2ex\noindent
(3)~{\it Gauge invariance.}
To preserve the gauge invariance of the theory
we require that $\langle{\cal O}\rangle_{\rm F}$ 
is a gauge-invariant function of the gauge field 
if $\cal O$ is a gauge-invariant product of the link variables
and the fermion fields.
In particular, the partition function $\langle 1\rangle_{\rm F}$
should be invariant and from our discussion in subsect.~3.5 
it is immediately clear that 
this condition will be fulfilled if
$$
  \drvstar{\mu}j_{\mu}(x)=\anomaly_L(x).
  \eqno\enum
$$
It is possible to prove that no further conditions arise
when one considers arbitrary products $\cal O$ of fields, 
i.e.~the gauge invariance of the theory is guaranteed if eq.~(4.2) holds. 
One of the consequences of this equation and
the integrability condition (3.12) is, incidentally, that 
the current $j_{\mu}(x)$ itself has to be gauge-invariant. 

\vskip2ex\noindent
(4)~{\it Lattice symmetries.}
In the continuum limit 
the imaginary part of the effective action transforms in a 
particular way under the space-time symmetries.
On the lattice 
one would like to preserve 
these symmetries as far as possible 
so as to reduce any remaining phase ambiguity
in the fermion integration measure.

Lattice translations, hyper-cubic rotations, reflections 
at the lattice planes and
charge conjugation will be referred to as 
the ``lattice symmetries" in the following.
We now demand that the measure term $\L_{\eta}$ transforms in the same way
under these symmetries as the imaginary part of the first term 
on the right-hand side of eq.~(3.8).
This is equivalent to requiring the current $j_{\mu}(x)$ 
to transform like the axial current
$$
  j^{5}_{\mu}(x)=\frac{1}{2}\left\{
  \psibar(x)\dirac{5}\dirac{\mu}U(x,\mu)\psi(x+\hat{\mu})+
  \psibar(x+\hat{\mu})\dirac{5}\dirac{\mu}U(x,\mu)^{-1}\psi(x)\right\}
  \eqno\enum
$$
in ordinary lattice gauge theories with Wilson-Dirac fermions.

\section 5. Statement of results

In the remainder of this paper we shall show
that fermion integration measures satisfying conditions (1)--(4)
exist in all topological sectors provided
$$
  \hbox{$N_{\rme}$ is even for all odd $\rme$},
  \eqno\enum
$$
where $N_{\rme}$ denotes the number of fermion
flavours with $|\rme_{\alpha}|=\rme$.
This includes the multiplet (2.14) and all cases with only even charges.
In the vacuum sector there is actually no restriction on the charge 
assignment apart from the anomaly cancellation condition
and it is currently not clear whether the constraint (5.1) 
reflects a fundamental limitation in finite volume or 
just a temporary technical difficulty.

For clarity the main steps of the construction  
are presented below in the form of three theorems
together with some key formulae.
All proofs are postponed to the later sections which 
should be consulted for full details.

\subsection 5.1 Reconstruction theorem

While the fermion integration measure is a relatively
complicated object, requiring the specification of 
a basis $v_j(x)$ of left-handed fields modulo measure preserving
basis transformations,
the associated current $j_{\mu}(x)$
is invariant under such transformations and is clearly much
more tractable. 
The following theorem says that the measure can be reconstructed
from the current under certain conditions.

\vskip2ex\noindent
{\bf Theorem 5.1.}{\sl\kern1ex
Suppose $j_{\mu}(x)$ is a given current
with the following properties.

\vskip1.2ex\noindent\kern1em\hbox{\vbox{\hsize=0.9\hsize
\noindent
(a)~$j_{\mu}(x)$ is defined for all admissible gauge fields
and depends smoothly on the link variables.

\vskip1ex\noindent
(b)~$j_{\mu}(x)$ is gauge-invariant and transforms as an axial vector current
under the lattice symmetries, as described in sect.~4.

\vskip1ex\noindent
(c)~The linear functional $\L_{\eta}=\sum_{x\in\Gamma}\eta_{\mu}(x)j_{\mu}(x)$
satisfies the integrability condition (3.12).

\vskip1ex\noindent
(d)~The anomalous conservation law $\drvstar{\mu}j_{\mu}(x)=\anomaly_L(x)$
holds.
}}

\vskip1ex\noindent
Then there exists a smooth fermion integration measure 
in the vacuum sector such that the associated
current coincides with $j_{\mu}(x)$. The same is true
in all other sectors if the charges satisfy the constraint (5.1).
In each case the measure is uniquely determined up to a constant phase factor.
}

\vskip2ex
We are thus left with the problem to find a local current $j_{\mu}(x)$ 
with the properties listed above.
Since the notion of locality which is being adopted here allows 
for exponentially decaying tails (with a fixed localization range in 
lattice units), the current can have 
non-local contributions that are of this order in the lattice size $L$.
In the following our strategy will be to provide an explicit
expression for the current in infinite volume and to prove 
that a solution in finite volume can be obtained by adding 
an exponentially small correction.

\subsection 5.2 Anomaly cancellation

Before proceeding with the construction of the current
it is however useful to discuss the significance 
of the anomaly cancellation condition (1.1) in the present framework.
For simplicity we consider the theory 
in infinite volume in this subsection.
The properties of the Dirac operator listed in appendix B then imply that 
the anomaly 
$$
  \anomaly(x)=-\frac{1}{2}\kern1.0pt\tr\left\{\dirac{5}TD(x,x)\right\}
  \eqno\enum
$$
is a gauge-invariant local field.
Moreover, using the Ginsparg-Wilson relation,
the anomaly can be shown to be a topological field satisfying
$$
  \sum_{x\in\gz^4}\,\delta_{\eta}\anomaly(x)=0
  \eqno\enum
$$
for any local deformation $\eta_{\mu}(x)$ of the gauge field.
It follows from this and a general theorem established in 
ref.~[\ref{LuscherII}] that
$$
  \anomaly(x)=
  \gamma\epsilon_{\mu\nu\rho\sigma}
  F_{\mu\nu}(x)F_{\rho\sigma}(x+\hat{\mu}+\hat{\nu})
  +\drvstar{\mu}k_{\mu}(x),
  \eqno\enum
$$
where $\gamma$ is a constant and $k_{\mu}(x)$ a gauge-invariant
local current. 

We now show that $\gamma=0$ by noting that
the Dirac operator is equal to 
the same analytic expression
for each fermion flavour $\alpha$, with
the link variables $U(x,\mu)$ replaced by $U(x,\mu)^{\rme_{\alpha}}$.
The field tensor scales with the charge and 
there is another power of the charge coming from the generator
$T$ in eq.~(5.2). 
The contribution to the constant $\gamma$ of the fermion with flavour $\alpha$ 
is hence proportional to $\rme_{\alpha}^3$ and after summing 
over all flavours one gets zero because of eq.~(1.1).

The anomaly thus cancels up to a divergence term.
At first sight one might think that this is not enough
to achieve the gauge invariance of the theory, but 
we only need to satisfy eq.~(4.2) for this and 
it is then conceivable that the gauge field dependence 
of the measure exactly compensates for the divergence term.
The important point to note here is 
that one would be unable to cancel the term proportional to $\gamma$
in this way. The construction of 
a fermion integration measure complying with conditions (1)--(4)
is hence only possible for anomaly-free fermion multiplets.

\subsection 5.3 Solution of the integrability condition in infinite volume

One of the technical advantages which one has in infinite volume is
that the gauge fields can be represented in a natural way 
through vector fields. The relevant lemma has been proved
in ref.~[\ref{LuscherII}] and is quoted here for the reader's convenience.

\proclaim Lemma 5.2.
Suppose $U(x,\mu)$ is an admissible gauge field on the infinite lattice.
Then there exists a vector field $A_{\mu}(x)$ such that 
$$
  \eqalignno{
  &U(x,\mu)=\rme^{iA_{\mu}(x)},
  \qquad
  |A_{\mu}(x)|\leq\pi\left(1+8\|x\|\right),
  &\enum\cr
  \noalign{\vskip2ex}
  &F_{\mu\nu}(x)=\drv{\mu} A_{\nu}(x)-\drv{\nu} A_{\mu}(x).
  &\enum\cr}
$$
Moreover, any other field with these 
properties is equal to $A_{\mu}(x)+\drv{\mu}\omega(x)$,
where the gauge function $\omega(x)$ takes values that are integer
multiples of $2\pi$.

\noindent
The idea is now to construct a solution of the 
integrability condition first in terms of the vector field.
So let us assume that $A_{\mu}(x)$ is any given field
representing an admissible gauge field $U(x,\mu)$ as in lemma 5.2.
The curve
$$
  U_t(x,\mu)=\rme^{itA_{\mu}(x)},
  \qquad
  0\leq t\leq1,
  \eqno\enum
$$
contracts this field to the classical vacuum 
configuration in such a way
that the field tensor remains bounded by $\epsilon$ for all $t$.
For any variation $\eta_{\mu}(x)$
of the gauge potential with compact support,
a linear functional $\Linfty_{\eta}$ may thus be defined through
$$
  \eqalignno{
  \Linfty_{\eta}&=i\int_0^1\rmd t\;
  \Tr\bigl\{\hat{P}_{-}
  \bigl[\partial_t\hat{P}_{-},\delta_{\eta}\hat{P}_{-}\bigr]\bigr\}+
  &\cr
  \noalign{\vskip2ex}
  &\phantom{=i}
  \int_0^1\rmd t\,\sum_{x\in\gz^4}
  \bigl\{\eta_{\mu}(x)\bar{k}_{\mu}(x)+
  A_{\mu}(x)\delta_{\eta}\bar{k}_{\mu}(x)\bigr\},
  &\enum\cr}
$$
where $\bar{k}_{\mu}(x)$ is any gauge-invariant local current,
which transforms as an axial vector field under the lattice symmetries 
and which satisfies $\drvstar{\mu}\bar{k}_{\mu}(x)=\anomaly(x)$.
An example of such a field is obtained by averaging
the current $k_{\mu}(x)$ introduced in subsect.~5.2 
over the lattice symmetries,
with the appropriate weights so as to project to the axial vector component.
Note that an explicit 
although very complicated expression for
$k_{\mu}(x)$ in terms of the first and second variations of the anomaly
has been derived in ref.~[\ref{LuscherII}].
The existence of a current $\bar{k}_{\mu}(x)$ 
with the required properties is thus guaranteed.

\vskip2.0ex\noindent
{\bf Theorem 5.3.}{\sl\kern1ex
The linear functional 
$\Linfty_{\eta}=\sum_{x\in\gz^4}\eta_{\mu}(x)\jinfty_{\mu}(x)$ 
defined above has the following properties.

\vskip1.2ex\noindent\kern1em\hbox{\vbox{\hsize=0.9\hsize
\noindent
(a)~$\Linfty_{\eta}$ is invariant under gauge transformations 
$A_{\mu}(x)\to A_{\mu}(x)+\drv{\mu}\omega(x)$, for arbitrary gauge
functions $\omega(x)$ that are polynomially bounded at infinity.

\vskip1ex\noindent
(b)~The current $\jinfty_{\mu}(x)$ is a local field, which depends smoothly
on the gauge field and 
which transforms as an axial vector current under the lattice symmetries.

\vskip1ex\noindent
(c)~$\Linfty_{\eta}$ is a solution of the integrability condition
$$
  \delta_{\eta}\Linfty_{\zeta}-\delta_{\zeta}\Linfty_{\eta}
  =i\,\Tr\bigl\{\hat{P}_{-}
  \bigl[\delta_{\eta}\hat{P}_{-},\delta_{\zeta}\hat{P}_{-}\bigr]\bigr\}
  \eqno\enum
$$
in infinite volume 
for all compactly supported variations
$\eta_{\mu}(x)$ and $\zeta_{\mu}(x)$.

\vskip1ex\noindent
(d)~The anomalous conservation law $\drvstar{\mu}\jinfty_{\mu}(x)=\anomaly(x)$
holds.
}}}

\vskip2.0ex
An important consequence of the gauge invariance of $\Linfty_{\eta}$ is that 
the current $\jinfty_{\mu}(x)$
may be considered to be a function of the gauge field $U(x,\mu)$
rather than the vector field $A_{\mu}(x)$, since the mapping
between the two is one-to-one modulo gauge transformations.
It can be shown that the locality,
differentiability and symmetry properties of the current are the same
independently of which point of view is adopted
[\ref{LuscherII}].

\subsection 5.4 Construction of the current $j_{\mu}(x)$ in finite volume

We now return to the theory in finite volume and first note that 
$\jinfty_{\mu}(x)$ becomes a gauge-invariant local field 
on the finite lattice
if attention is restricted to periodic gauge fields. 
As asserted by the following theorem, this current
has all the required properties up to exponentially 
small finite-lattice corrections.

\vskip2ex\noindent
{\bf Theorem 5.4.}{\sl\kern0.8ex
If the lattice is sufficiently large
compared to the localization range~$\varrho$ of the Dirac operator,
say $L/\varrho\geq n$,
there exists a current $j_{\mu}(x)$ which satisfies 
$$
  |j_{\mu}(x)-\jinfty_{\mu}(x)|
  \leq \kappa L^{\nu}\kern0.5pt\rme^{-L/\varrho}
  \eqno\enum
$$
and which fulfils conditions (a)--(d) of theorem 5.1.
The bound (5.10) holds uniformly in the gauge field, 
i.e.~the constants $n$, $\kappa$ and $\nu$ are independent of the field.
}

\vskip2ex\noindent
Together with theorem 5.1 this result implies that fermion integration
measures satisfying conditions (1)--(4) exist on large lattices.
Note that the difference between $j_{\mu}(x)$ and $\jinfty_{\mu}(x)$ 
vanishes exponentially
in the continuum limit, because 
$\varrho$ is a fixed number in lattice units
while $L$ is a physical length scale. The detailed form of these
corrections is hence of little interest. 

The theorems quoted in this section are not easy to prove.
Most of the difficulties can be traced back to the fact that the 
space of admissible gauge fields is topologically non-trivial
in finite volume.
Differential geometry and the theory of fibre bundles
are the adequate tools to deal with this problem 
and the reader who wishes to go through the details 
in sects.~7--11 will be assumed 
to be familiar with the relevant mathematical terminology.

\section 6. Proof of theorem 5.3

We first remark that the projector
$\hat{P}_{-}$ has the same 
locality properties as the Dirac operator.
In particular, the kernel of $\delta_{\eta}\hat{P}_{-}$
falls off exponentially away
from the support of $\eta_{\mu}(x)$ and the trace in eq.~(5.8)
is hence rapidly convergent in position space.
One of the consequences of this technical observation is 
that $\Linfty_{\eta}$ is a well-defined and smooth function 
of the gauge potential $A_{\mu}(x)$.
We now establish the other properties of $\Linfty_{\eta}$
in the order stated in the theorem.

\vskip2ex
\noindent
(a)~{\it Gauge invariance.}
Taking the gauge covariance of the projector $\hat{P}_{-}$
and the gauge invariance of the current $\bar{k}_{\mu}(x)$ into account,
it is easy to show that the change of
$\Linfty_{\eta}$ under gauge transformations 
$A_{\mu}(x)\to A_{\mu}(x)+\drv{\mu}\omega(x)$ is given by
$$
  \int_0^1\rmd t\,
  \Tr\bigl\{\hat{P}_{-}
  \bigl[[\omega T,\hat{P}_{-}],\delta_{\eta}\hat{P}_{-}\bigr]\bigr\}+
  \int_0^1\rmd t\sum_{x\in\gz^4}\drv{\mu}\omega(x)\delta_{\eta}\bar{k}_{\mu}(x).
  \eqno\enum
$$
Expanding the commutators and using the identity
$$
  \hat{P}_{-}\delta_{\eta}\hat{P}_{-}\hat{P}_{-}=0,
  \eqno\enum
$$
the first term can be rewritten in the form
$$
  \int_0^1\rmd t\sum_{x\in\gz^4}\omega(x)\delta_{\eta}\anomaly(x),
  \eqno\enum
$$
where $\anomaly(x)$ denotes the anomaly in infinite volume.
Recalling $\drvstar{\mu}\bar{k}_{\mu}(x)=\anomaly(x)$
and performing a partial summation it is now clear  
that the two terms in eq.~(6.1) cancel each other.

\vskip2ex
\noindent
(b)~{\it Locality and symmetry properties of $\jinfty_{\mu}(x)$.}
From what has been said at the beginning of this section, 
and since $\bar{k}_{\mu}(x)$ is 
a smooth local function of the gauge field,
it is evident that the same is true for $\jinfty_{\mu}(x)$.
Moreover under the lattice symmetries
it transforms as an axial vector field.
To prove this for space-time reflections one has to take 
into account that 
$$
  \Tr\bigl\{\hat{P}_{+}
  \bigl[\partial_t\hat{P}_{+},\delta_{\eta}\hat{P}_{+}\bigr]\bigr\}
  =
  -\Tr\bigl\{\hat{P}_{-}
  \bigl[\partial_t\hat{P}_{-},\delta_{\eta}\hat{P}_{-}\bigr]\bigr\}.
  \eqno\enum
$$
In all other cases the transformations commute with the projection
to the left-handed fields and the covariance of the current is 
deduced straightforwardly. 

\vskip2ex
\noindent
(c)~{\it Integrability condition.} 
Starting from the definition (5.8) of $\Linfty_{\eta}$,
one quickly finds that the second term does not contribute
to the curvature $\delta_{\eta}\Linfty_{\zeta}-\delta_{\zeta}\Linfty_{\eta}$.
Another simple observation is that all terms of the form
$$
  \Tr\bigl\{
  \delta_1\hat{P}_{-}\,\delta_2\hat{P}_{-}\,\delta_3\hat{P}_{-}
  \bigr\}
  \eqno\enum
$$
can be shown to vanish by inserting $1=(\hatdirac{5})^2$
and using the fact that $\hatdirac{5}$ anti-commutes with 
any variation of the projector $\hat{P}_{-}$.
Taking this into account, there are only two terms which 
contribute to the curvature,
$$
  i\int_0^1\rmd t\,\Tr\bigl\{
  \hat{P}_{-}
  \bigl[\delta_{\eta}\partial_t\hat{P}_{-},\delta_{\zeta}\hat{P}_{-}\bigr]-
  \hat{P}_{-}
  \bigl[\delta_{\zeta}\partial_t\hat{P}_{-},\delta_{\eta}\hat{P}_{-}\bigr]
  \bigr\},
  \eqno\enum
$$
and these may be rewritten in the form
$$
  i\int_0^1\rmd t\,\partial_t\Tr\bigl\{
  \hat{P}_{-}
  \bigl[\delta_{\eta}\hat{P}_{-},\delta_{\zeta}\hat{P}_{-}\bigr]
  \bigr\}.
  \eqno\enum
$$
After integration one then ends up with eq.~(5.9)
since the contribution from the lower end of the integration range
is equal to zero.

\vskip2ex
\noindent
(d)~{\it Anomalous conservation law.}
Setting $\eta_{\mu}(x)=-\drv{\mu}\omega(x)$
(where $\omega(x)$ is any lattice function with compact support)
the left-hand side of eq.~(5.8) becomes
$$
  \sum_{x\in\gz^4}\omega(x)\drvstar{\mu}\jinfty_{\mu}(x).
  \eqno\enum
$$
On the other side we insert the identities
$$
  \delta_{\eta}\hat{P}_{-}=it\bigl[\omega T,\hat{P}_{-}\bigr],
  \qquad
  \delta_{\eta}\bar{k}_{\mu}(x)=0,
  \eqno\enum
$$
and in a few steps obtain a sum of two terms, 
$$
  -\int_0^1\rmd t\,t\,\Tr\bigl\{\omega T\partial_t\hat{P}_{-}\bigr\}
  +\int_0^1\rmd t \sum_{x\in\gz^4}\omega(x)\anomaly(x).
  \eqno\enum
$$
Expressing the trace in the first term through the anomaly (5.2),
the terms nearly cancel after a partial integration
and one is left with the contribution 
$$
  \sum_{x\in\gz^4}\omega(x)\left.\anomaly(x)\right|_{t=1}
  \eqno\enum
$$
from the upper end of the integration range.
Comparing with eq.~(6.8) this shows that the divergence of 
$\jinfty_{\mu}(x)$ is equal to the anomaly and thus completes the proof 
of the theorem.

\section 7. Topology of the field space in finite volume

We now begin with the detailed discussion of the theory in finite
volume and first determine the structure of the 
space of admissible gauge fields. As will be explained in the next section,
the existence of smooth fermion integration
measures depends on whether a certain U(1) bundle over this space
is trivial. 
Evidently, to be able to address this problem,
one needs to know the topology of the base manifold.

\subsection 7.1 Preliminaries

In the following the lattice is 
assumed to be finite with periodic boundary conditions as specified
in sect.~2. The space of admissible gauge fields
is denoted by $\gfields$ and the gauge group $\group$ is taken 
to be the subset of gauge transformations $\Lambda(x)$
satisfying $\Lambda(x)=1$ at $x=0\mod L$.

For any given gauge field $U(x,\mu)$, the Wilson lines winding 
around the lattice along the coordinate axes are defined by
$$
  W_{\mu}(x)=\prod_{s=0}^{L-1}U(x+s\hat{\mu},\mu).
  \eqno\enum
$$
They are gauge-invariant, but cannot be expressed through the 
field tensor $F_{\mu\nu}(x)$ and thus carry independent information
on the gauge field. 

\proclaim Lemma 7.1.
Any two admissible fields
$U(x,\mu)$ and $\widetilde{U}(x,\mu)$ satisfying 
$$
  F_{\mu\nu}(x)=\widetilde{F}_{\mu\nu}(x)
  \quad\hbox{and}\quad
  W_{\mu}(x)=\widetilde{W}_{\mu}(x)
  \eqno\enum
$$
are gauge equivalent.

\proof
If we introduce a new field through
$$
  V(x,\mu)=U(x,\mu)\widetilde{U}(x,\mu)^{-1},
  \eqno\enum
$$
it is obvious that the associated plaquette loops 
and Wilson lines are all equal to~$1$.
The product $\Lambda(x)$ of the link variables $V(x,\mu)$ along any lattice 
path from $x$ to the 
origin $x=0$ is hence independent of the chosen path and periodic in $x$.
In other words, $\Lambda(x)$ is an element of the gauge group $\group$
which transforms $V(x,\mu)$ to $1$ and thus
$U(x,\mu)$ to $\widetilde{U}(x,\mu)$.
\endproof

\vskip2ex
The subspace $\cfields$ 
of all admissible gauge fields with vanishing field tensor 
contains the pure gauge configurations, but there are also
non-trivial configurations with Wilson lines different from $1$.
It is straightforward to show, however, that the Wilson lines $W_{\mu}(x)$
do not depend on $x$. The gauge-invariant content of such fields
is hence encoded in the constant phase factors $w_{\mu}=W_{\mu}(x)$.

\proclaim Lemma 7.2.
The gauge fields with vanishing field tensor are of the form
$$
  U(x,\mu)=\Lambda(x)U_{[w]}(x,\mu)\Lambda(x+\hat{\mu})^{-1},
  \eqno\enum
$$
where $\Lambda(x)$ is an element of\/ $\group$ and
the field $U_{[w]}(x,\mu)$ is defined by 
$$
  U_{[w]}(x,\mu)=\cases{w_{\mu} & if $x_{\mu}=0\mod L$, \cr
                    \noalign{\vskip1ex}
                    1       & otherwise,\cr}
  \eqno\enum
$$
for any given set of phase factors $w_{\mu}\in\rmU(1)$.
Moreover the representation (7.4) is unique and
establishes the isomorphism\/ $\cfields\cong\rmU(1)^4\times\group$.

\proof
From the definition (7.5) it is obvious that $U_{[w]}(x,\mu)$ is 
a gauge field with vanishing field tensor and Wilson lines $W_{\mu}(x)$ equal
to $w_{\mu}$. According to lemma 7.1, any other field $U(x,\mu)$ with these
properties is gauge equivalent to $U_{[w]}(x,\mu)$. 
This proves eq.~(7.4) and it is now 
also evident that $w_{\mu}$ and $\Lambda(x)$ are uniquely 
determined by the gauge field. 
\endproof

\vskip1ex

\subsection 7.2 Flux sectors

As already mentioned in subsect.~2.2, the field space $\gfields$
is a union of dis\-con\-nected subspaces $\gfields[m]$
labelled by the magnetic flux quantum numbers $m_{\mu\nu}$. 
We now prove this and provide some further information
on the flux sectors.

\proclaim Lemma 7.3.
Let $U(x,\mu)$ be an admissible gauge field and 
define the associated magnetic flux $\phi_{\mu\nu}(x)$
through eq.~(2.10). Then there exists an anti-symmetric integer tensor
$m_{\mu\nu}$ such that 
$\phi_{\mu\nu}(x)=2\pi m_{\mu\nu}$ for all $x$. 
  
\proof
If we define a vector potential $a_{\mu}(x)$ through
$$
  a_{\mu}(x)={1\over i}\ln U(x,\mu),\qquad
  -\pi<a_{\mu}(x)\leq \pi,
  \eqno\enum
$$
it is straightforward to show that 
$$
  F_{\mu\nu}(x)=\drv{\mu} a_{\nu}(x)-\drv{\nu} a_{\mu}(x)+2\pi z_{\mu\nu}(x),
  \eqno\enum
$$
where $z_{\mu\nu}(x)$ takes integer values. Only the second term in 
this equation contributes to the magnetic flux which
is hence an integer multiple of $2\pi$.

The periodicity of the field tensor implies that  
$\phi_{\mu\nu}(x)$ is independent of the coordinates $x_{\mu}$
and $x_{\nu}$. 
To prove that the flux is also independent 
of the complementary components of $x$, we note that
$$
  \epsilon_{\mu\nu\rho\sigma}\drv{\nu}F_{\rho\sigma}(x)=0.
  \eqno\enum
$$
This is a straightforward consequence of lemma 5.2 and 
particularly of eq.~(5.6).
Using the periodicity of the field tensor again and partial summations,
the change $\drv{\rho}\phi_{\mu\nu}(x)$ of the flux in any direction 
orthogonal to the $(\mu,\nu)$--plane is then easily shown to vanish.
\endproof

\vskip2ex
As long as only admissible fields are considered,
the field tensor is a continuous function of the link variables and the 
magnetic flux quantum numbers consequently 
cannot change under continuous deformations of the field.
The sectors $\gfields[m]$ of all fields with 
a given set of flux quantum numbers $m_{\mu\nu}$ are thus 
disconnected from each other.
There are at most a finite
number of sectors since
$$
  |m_{\mu\nu}|<{\epsilon\over2\pi}L^2
  \eqno\enum
$$
as one may easily prove by combining eqs.~(2.9) and (2.10).
Conversely if $m_{\mu\nu}$ is any prescribed, anti-symmetric integer tensor
satisfying this bound,  
there exist admissible fields with these flux quantum numbers.
An example of such a field is
$$
  V_{[m]}(x,\mu)=
  \exp\,\biggl\{-{2\pi i\over L^2}
  \biggl[
  L\delta_{\tilde{x}_{\mu},L-1}
  \sum_{\nu>\mu}\,m_{\mu\nu}\tilde{x}_{\nu}+
  \sum_{\nu<\mu}\,m_{\mu\nu}\tilde{x}_{\nu}
  \biggr]\biggr\},
  \eqno\enum
$$
where the abbreviation $\tilde{x}_{\mu}=x_{\mu}\mod L$ has been used.
This field is periodic and can be shown to have constant field tensor equal to 
$2\pi m_{\mu\nu}/L^2$.

\subsection 7.3 Topology of\/ $\gfields[m]$

We now determine the structure of the flux sector $\gfields[m]$
for any given set of flux quantum numbers $m_{\mu\nu}$.
As will be shown below, one of the factors of this manifold 
consists of the space 
$\vfields[m]$ of all periodic vector potentials $\Atrans_{\mu}(x)$
satisfying
$$
  \eqalignno{
  &\drvstar{\mu}\Atrans_{\mu}(x)=0,
  \qquad
  \sum_{x\in\Gamma}\,\Atrans_{\mu}(x)=0,
  &\enum\cr
  \noalign{\vskip2ex}
  &\left|\drv{\mu}\Atrans_{\nu}(x)-\drv{\nu}\Atrans_{\mu}(x)
  +2\pi m_{\mu\nu}/L^2\right|<\epsilon.
  &\enum\cr}
$$
The index ``T" reminds us that these fields are transverse and also 
serves to distinguish them from the vector potential $A_{\mu}(x)$
which has been introduced in sect.~5. 
Note that $\vfields[m]$ is a convex space.
In particular, it is contractible and thus topologically trivial.

\proclaim Lemma 7.4. The fields $U(x,\mu)$ in the sector $\gfields[m]$
are of the form
$$
  U(x,\mu)=V_{[m]}(x,\mu)\ring{U}(x,\mu)\kern0.5pt\rme^{i\atrans_{\mu}(x)},
  \eqno\enum
$$
where $\ring{U}(x,\mu)$ has vanishing field tensor and $\Atrans_{\mu}(x)$
is an element of\/ $\vfields[m]$. Moreover this representation
is unique and establishes the isomorphism 
$\gfields[m]\cong\gfields_0\times\vfields[m]$.

\proof
We first prove the 
uniqueness of the representation (7.13) by noting that
the field tensor of $U(x,\mu)$ is given by
$$
  F_{\mu\nu}(x)=
  \drv{\mu}\Atrans_{\nu}(x)-\drv{\nu}\Atrans_{\mu}(x)+2\pi m_{\mu\nu}/L^2.
  \eqno\enum
$$
Together with the constraints (7.11) this equation implies that
$$
  \Atrans_{\mu}(x)=
  \sum_{y\in\Gamma}G_L(x-y)\drvstar{\lambda}F_{\lambda\mu}(y),
  \eqno\enum
$$
where $G_L(z)$ denotes the Green function of the lattice laplacian,
$$
  \drvstar{\mu}\drv{\mu}G_L(z)=\delta_{\tilde{z}0}-L^{-4},
  \qquad
  G_L(z+L\hat{\mu})=G_L(z),
  \qquad
  \sum_{z\in\Gamma}G_L(z)=0.
  \eqno\enum
$$
In particular, the transverse field is uniquely determined
and so are the other factors in eq.~(7.13).

To show that any given admissible field
$U(x,\mu)$ with field tensor $F_{\mu\nu}(x)$
and flux quantum numbers $m_{\mu\nu}$ can be represented in this
way, we turn the argument around and define $\Atrans_{\mu}(x)$ 
through eq.~(7.15). From the properties of the Green function
it is then clear that this field satisfies eq.~(7.11).
Moreover, using eq.~(7.8) (which holds for any admissible field) and 
the fact that the zero-momentum component of $F_{\mu\nu}(x)$ is 
proportional to $m_{\mu\nu}$, it is straightforward to establish 
eq.~(7.14). 
In particular, $\Atrans_{\mu}(x)$ is contained in $\vfields[m]$ and 
$$
  \ring{U}(x,\mu)=V_{[m]}(x,\mu)^{-1}U(x,\mu)\kern0.5pt\rme^{-i\atrans_{\mu}(x)}
  \eqno\enum
$$
has vanishing field tensor.
\endproof

\vskip2ex
Taken together the results obtained in this section imply that 
$$
  \gfields[m]\cong\rmU(1)^4\times\group\times\vfields[m].
  \eqno\enum
$$
Since $\group$ is a product of U(1) factors, the sectors 
$\gfields[m]$ are thus either empty or equal to a multi-dimensional
torus times a contractible space.

\section 8. Fermion integration measures and $\rmU(1)$ 
bundles over $\gfields$

One of the conditions on the fermion integration measure listed
in sect.~4 is that the fermion expectation values 
$\langle{\cal O}\rangle_{\rm F}$ should be smooth functions of the 
gauge field.
In this section the implications of this requirement
for the basis vectors $v_j(x)$ are worked out and
we shall then be able to reformulate the condition in geometrical terms 
which will later allow us to make use of some known results of 
the theory of fibre bundles.

\subsection 8.1 Smooth measures

Evidently the differentiability of the fermion integrals
$\langle{\cal O}\rangle_{\rm F}$ will be guaranteed 
if the basis vectors
$v_j(x)$ are globally defined and smooth, but since the space of admissible
gauge fields is topologically non-trivial one may be unable to 
find such a basis. 

We can, however, cover the 
space of admissible fields with open contractible patches $X_a$,
labelled by an index $a$, and choose a smooth basis $v^a_j(x)$ 
on each of these patches. Since the
projector to the left-handed fields depends smoothly on the gauge field
and since the field manifold $\gfields$ is smooth, this is always 
possible.
On the intersection $X_a\cap X_b$
of any two patches, we then have two bases
which are related to each other by a unitary transformation 
$$
  v_j^b(x)=\sum_k\,v_k^a(x)\, \trfun(a\to b)_{kj}
  \eqno\enum
$$
as in eq.~(2.26).
The transition matrices $\trfun(a\to b)$ satisfy the cocycle condition
$$
  \trfun(a\to c)=\trfun(a\to b)\trfun(b\to c)
  \quad\hbox{on}\quad
  X_a\cap X_b\cap X_c
  \eqno\enum
$$
and thus define a unitary 
principal bundle over $\gfields$.

The fermion integration measure
changes by a phase factor equal to the determinant of the 
transition matrix if one passes from one basis to another.
We now only need to make sure that 
$$
  \det \trfun(a\to b)=1
  \eqno\enum
$$
on the intersections $X_a\cap X_b$.
The integration measure is then independent
of the patch and the fermion integrals 
$\langle{\cal O}\rangle_{\rm F}$ are thus globally defined and smooth.

If $v_j^a(x)$ is any patched basis, not necessarily satisfying
eq.~(8.3), it is clear from the above that 
the phase factors 
$$
  g_{ab}=\det\trfun(a\to b)
  \eqno\enum
$$
define a U(1) bundle over $\gfields$.
Moreover, under a change of basis, these phase factors 
transform according to 
$$
  g_{ab}\to h_ag_{ab}{h_b}^{\kern-4pt\raise1pt\hbox{$\scriptstyle -1$}}
  \quad\hbox{on}\quad X_a\cap X_b,
  \eqno\enum
$$
where $h_a$ is the determinant of the transformation matrix on patch $X_a$.
Different choices of the basis vectors thus give rise to isomorphic bundles.
The converse is also true since for any set of smooth 
phase factors $h_a$ one can always find a corresponding basis transformation.

It should now be evident that smooth fermion integration 
measures exist if (and only if) this bundle is trivial.
Whether this is the case depends on the properties of 
the projector to the left-handed fields and the base manifold $\gfields$.
If the bundle is non-trivial one has an anomaly
and it is then not possible to construct fermion integration measures
satisfying condition (1)
\footnote{$\dagger$}{\footnotefont
The discussion in this section is closely related to 
the recent work of Neuberger [\ref{OverlapAnomaly}]
on the gauge anomaly in the overlap formalism.
An explicit example is given there demonstrating the presence 
of a non-integrable phase if the fermion multiplet is 
not anomaly-free}.

\subsection 8.2 Geometrical interpretation of the measure term $\L_{\eta}$

For any given basis $v^a_j(x)$
the measure term is defined through
$$
  \L^a_{\eta}=i\sum_j\,(v^a_j,\delta_{\eta}v^a_j)
  \eqno\enum
$$ 
on patch $X_a$.
The transition rule for passing from one patch to another is
$$
  \L^a_{\eta}=\L^b_{\eta}
  -ig_{ab}^{-1}
  \delta_{\eta}g_{ab}^{\phantom{-1}}
  \quad\hbox{on}\quad X_a\cap X_b
  \eqno\enum
$$
and $\L^a_{\eta}$ is thus
a connection on the U(1) bundle constructed
above. Note that $\L^a_{\eta}$ is independent of the patch if 
the smoothness condition (8.3) is fulfilled. The associated 
current $j_{\mu}(x)$ is then a globally defined smooth function
of the gauge field.

As we have previously remarked, the curvature (3.12) 
of the measure term is invariant under basis transformations.
Evidently the curvature is just the field strength of 
the connection $\L^a_{\eta}$. 
The Wilson lines constructed from $\L^a_{\eta}$ winding around
the base manifold $\gfields$ in a particular direction
are further invariants that carry important
information on the measure term.

To make this completely clear let us consider a closed curve
$$
  U_t(x,\mu), 
  \qquad
  0\leq t\leq 2\pi,
  \eqno\enum
$$
of admissible fields. If we temporarily assume that the smoothness condition
(8.3) is satisfied, 
the Wilson line associated with the curve is given by
$$
  \Wline=\exp\biggl\{i\int_0^{2\pi}\rmd t\,\L_{\eta}\biggr\},
  \qquad
  \eta_{\mu}(x)=-iU_t(x,\mu)^{-1}\partial_tU_t(x,\mu).
  \eqno\enum
$$
The patch label has been dropped here, because the measure term
does not depend on it if (8.3) holds. Note that the 
variation $\eta_{\mu}(x)$ may be regarded as the tangential vector
along the curve. In particular, 
$$
  \L_{\eta}=i\sum_j\,(v^a_j,\partial_tv^a_j),
  \eqno\enum
$$
and it is then easy to check that the Wilson line does not depend on
the choice of basis. As shown by the following lemma, 
$\Wline$ may in fact be expressed directly through 
the projector to the left-handed fields.

\proclaim Lemma 8.1.
The Wilson line defined above is given by
$$
  \Wline=\lim_{n\to\infty}\det\bigl\{
  1-P_{t_0}+P_{t_n}P_{t_{n-1}}\ldots P_{t_0}\bigr\},
  \qquad
  t_k=2\pi k/n,
  \eqno\enum
$$
where $P_t$ is equal to the projector $\hat{P}_{-}$ along the curve (8.8).

\proof
Since we are considering a closed curve, we have $P_{t_n}=P_{t_0}$ and  
the determinant on the right-hand side of eq.~(8.11) is hence equal
to the determinant of the product $P_{t_n}P_{t_{n-1}}\ldots P_{t_0}$ in the 
subspace of left-handed fields at $t=0$.
To compute the determinant, we insert the representation
$$
  P_{t,L}(x,y)=\sum_j v^a_j(x)\otimes v^a_j(y)^{\ast}
  \eqno\enum
$$
for the kernels of the projectors $P_t$ in position space. 
One then obtains a product of matrices with entries
that are the scalar products of the basis vectors at subsequent
values of $t$. At large $n$ these matrices may be expanded according to
$$
  \delta_{lj}-(2\pi/n)(v^a_l,\partial_t v^a_j)_{t=t_k}+\rmO(1/n^2)
  \eqno\enum
$$
and for the determinant the expression 
$$
  \exp\Bigl\{i(2\pi/n)\sum_{k=0}^{n-1}\,(\L_{\eta})_{t=t_k}+\rmO(1/n)
  \Bigr\}
  \eqno\enum
$$
is thus obtained, 
which converges to $\Wline$ in the limit $n\to\infty$.
\endproof

\section 9. Abelian gauge fields on the $n$-dimensional torus

In each topological sector the submanifold of admissible fields 
where the transverse field $\Atrans_{\mu}(x)$ vanishes is an
$n$-dimensional torus
$$
  T^n=\rmU(1)\times\rmU(1)\times\ldots\times\rm U(1)
  \qquad
  \hbox{($n$ factors)}.
  \eqno\enum
$$
Since $\vfields[m]$ is contractible, a well-known theorem on fibre bundles
may be invoked which says that the U(1) bundle constructed in the 
preceding section is trivial if its restriction to $T^n$ is trivial.
U(1) bundles over $T^n$ can be completely classified and it is now helpful
to discuss this and the gauge fields that live on them in some detail.
The results quoted below are generally valid and do not 
refer to any particular properties of the bundle other than those
specified in the following paragraphs.
We shall then return to the case of interest in sects.~10 and 11.

\subsection 9.1\/ $\rmU(1)$ bundles over $T^n$

The points $u$ of the torus (9.1) may be locally parametrized through
$$
  u=(\rme^{it_1},\ldots,\rme^{it_n}),
  \eqno\enum
$$
where the coordinates $(t_1,\ldots,t_n)$ range in some small contractible
region in $\rz^n$. A particular choice of this region defines a 
coordinate patch $X_a$ on $T^n$ and the set of all these patches 
provides an atlas for this manifold. 

U(1) bundles over $T^n$
may be defined by specifying a set of smooth
transition functions $g_{ab}\in\rmU(1)$ 
on $X_a\cap X_b$ such that the cocycle condition
$$
  g_{ac}=g_{ab}g_{bc}
  \quad\hbox{on}\quad
  X_a\cap X_b\cap X_c
  \eqno\enum
$$
is satisfied. 
Two bundles with transition functions $g_{ab}$ 
and $\tilde{g}_{ab}$ are isomorphic if
$$
  \tilde{g}_{ab}=h_ag_{ab}{h_b}^{\kern-4pt\raise1pt\hbox{$\scriptstyle -1$}}
  \eqno\enum
$$
for some smooth gauge transformation functions $h_a\in\rmU(1)$ on $X_a$.
Isomorphic bundles can be continuously deformed into each other and
are thus topologically indistinguishable. In particular, any bundle 
which is isomorphic
to the bundle with transition functions $g_{ab}=1$ 
is referred to as trivial.

\subsection 9.2 Gauge fields and topological classification of\/ 
                $\rmU(1)$ bundles

A gauge field on a given U(1) bundle consists of 
a set of locally defined smooth vector fields $\Bfield^a_k$ 
such that 
$$
  \Bfield^a_k=\Bfield^b_k
  -ig_{ab}^{-1}
  \partial_kg_{ab}^{\phantom{-1}}
  \quad\hbox{on}\quad X_a\cap X_b
  \eqno\enum
$$
(the index $k$ and the derivative $\partial_k$ refer to the coordinates
$t_1,\ldots,t_n$). 
The associated field tensor
$\Gfield_{kl}=\partial_k\Bfield^a_l-\partial_l\Bfield^a_k$ is invariant
under these transformations and is thus independent of the patch label.
It can be shown that gauge fields exist on any bundle and
the following result is also well-known.

\proclaim Lemma 9.1.
The magnetic flux 
$$
  \frak I_{kl}=\int_0^{2\pi}\rmd t_k \rmd t_l\,\Gfield_{kl}
  \qquad
  \hbox{(no sum over $k$ and $l$)}
  \eqno\enum
$$
through the $(k,l)$--planes is quantized in units of\/ $2\pi$
and only depends on the underlying bundle. Moreover
any two bundles with the same flux quantum numbers
are isomorphic to each other.

\noindent
A given bundle is hence trivial if 
the integrals $\frak I_{kl}$ are equal to zero
for some particular gauge field. It is in fact sufficient to show that
$|\frak I_{kl}|<2\pi$ since the magnetic flux is quantized.

\subsection 9.3 Gauge fields on the trivial bundle

Gauge fields on the bundle with transition functions $g_{ab}=1$ 
may be represented by 
smooth periodic vector fields $\Bfield_k(t)$ on $\rz^n$
with period $2\pi$.
Such fields are not uniquely determined by the associated field tensor,
but one can always find a periodic gauge potential 
for a given field tensor if 
a few obvious conditions are fulfilled. 
The following lemma provides a particular solution of this problem.

\proclaim Lemma 9.2.
Suppose $\Gfield_{kl}(t)$ is a smooth periodic tensor field 
satisfying 
$$
  \Gfield_{kl}=-\Gfield_{lk},
  \qquad
  \partial_k\Gfield_{lj}+
  \partial_l\Gfield_{jk}+
  \partial_j\Gfield_{kl}=0.
  \eqno\enum
$$
If the associated magnetic fluxes (9.6) vanish, there exists
a smooth periodic vector field\/ $\Bfield_k(t)$ such that\/
$\Gfield_{kl}=\partial_k\Bfield_l-\partial_l\Bfield_k$ and 
$$
  \left|\Bfield_k(t)\right|\leq\pi(n-1)
  \sup_{r,k,l}\left|\Gfield_{kl}(r)\right|.
  \eqno\enum
$$

\proof
If $n=1$ there is nothing to prove
since $\Bfield_k(t)=0$ is a possible choice for the gauge field.
Now let us assume that the lemma has been established in dimension $n-1$
and that $\Gfield_{kl}(t)$ is a given tensor field in $n$
dimensions with the required properties. Evidently, 
when restricted to the hyper-plane $t_n=0$, this field
satisfies the premises of the lemma in $n-1$ dimensions 
and we may conclude that there exists a 
periodic vector field $\Bfield^{\circ}_k(t)$ 
depending on $t_1,\ldots,t_{n-1}$ such that
$$
  \left.\Gfield_{kl}(t)\right|_{t_n=0}=
  \partial_k\Bfield^{\circ}_l(t)-\partial_l\Bfield^{\circ}_k(t) 
  \quad\hbox{for all}\quad k,l<n.
  \eqno\enum
$$
In the following it will be convenient to consider
$\Bfield^{\circ}_k(t)$ to be a field on 
$\rz^n$ which is independent of $t_n$.

We now introduce the field
$$
  \bfield_k(t)=-\int_0^{2\pi}{\rmd r_n\over2\pi}\,\Gfield_{nk}(r),
  \qquad r=(t_1,\ldots,t_{n-1},r_n),
  \eqno\enum
$$
which is also independent of $t_n$ and periodic.
Using the properties of the field tensor,
it is easy  to show that $\partial_k\bfield_l-\partial_l\bfield_k=0$
and the line integral
$$
  \Bfield_n(t)=\int_0^t\rmd r_k\kern0.5pt \bfield_k(r)
  \eqno\enum
$$
is hence independent of the integration path.
Note that $\Bfield_n(t)$ is periodic in all coordinates $t_k$ since 
the flux integrals $\frak I_{nk}$ vanish.

Next we define the components of the vector field with 
index $k<n$ through
$$
  \Bfield_k(t)=
  \int_0^{t_n}\rmd r_n \Gfield_{nk}(r)+t_n\bfield_k(t)+\Bfield^{\circ}_k(t),
  \eqno\enum
$$
where $r$ is as in eq.~(9.10). This field is periodic
and it is evident that 
$$
  \partial_n\Bfield_k(t)=\Gfield_{nk}(t)+\bfield_k(t).
  \eqno\enum
$$
Together with eq.~(9.11) (which implies $\partial_k\Bfield_n=\bfield_k$),
this proves that 
$\Gfield_{nk}=\partial_n\Bfield_k-\partial_k\Bfield_n$
and it is easy to check that the other components of the field tensor 
are also correctly obtained.

To show that 
the so constructed gauge field satisfies the bound (9.8),
one proceeds inductively, assuming the bound holds 
for the field $\Bfield^{\circ}_{k}(t)$
with $n$ replaced by $n-1$. Straightforward estimates
of the right-hand sides of eqs.~(9.10)--(9.12),
taking the periodicity of the fields into account, then lead to the 
desired bound. 
\endproof

\vskip2ex
Another result on which we shall rely later is that 
the gauge-invariant content of a given gauge field $\Bfield_k(t)$
is completely determined by the associated field tensor
and the Wilson lines
$$
  \Wline_{k}(t)=\exp\biggl\{i\int_0^{2\pi}\!\rmd s
  \left.\Bfield_k(t)\right|_{t_k\to t_k+s}\biggr\}
  \eqno\enum
$$
winding around the torus.
This is a well-known result and we simply quote

\proclaim Lemma 9.3.
Any two smooth periodic gauge fields\/ $\Bfield_k(t)$ and\/ 
$\widetilde{\Bfield}_k(t)$ with the same
field tensors and the same Wilson lines are related 
to each other by a gauge transformation,
$$
  \widetilde{\Bfield}_k(t)=\Bfield_k(t)-ih(t)^{-1}\partial_kh(t),
  \eqno\enum
$$
where $h(t)\in\rmU(1)$ is a smooth periodic function of the 
coordinates $t_1,\ldots,t_n$.

\noindent
Note that the Wilson lines coincide at all $t$ if 
do at $t=0$ and if the field tensors of the two fields
are the same, since
$$
  \Wline_{k}(t)=\Wline_{k}(0)
  \exp\biggl\{i\int_0^t\rmd r_l\int_0^{2\pi}\!\rmd s
  \left.\Gfield_{lk}(r)\right|_{r_k\to r_k+s}
  \biggr\},
  \eqno\enum
$$
where the line integral from $0$ to $t$ is taken along an arbitrary path.

\section 10. Proof of theorem 5.1

We first show that smooth measures exist and shall then apply
a basis transformation to match $j_{\mu}(x)$ with the current
$\tilde{\jmath}_{\mu}(x)$ derived from the measure.

\subsection 10.1 Existence of a smooth measure

As discussed in sect.~8, we can always choose a patched basis 
$v_j^a(x)$ of left-handed fields which is locally smooth.
We now prove that the associated U(1) bundle is trivial.
Since the space $\vfields[m]$ of transverse vector fields is contractible,
it suffices to consider the bundle over the 
submanifold of admissible fields with $\Atrans_{\mu}(x)=0$.
Our task is then to show that the connection
$$
  \Bfield_k^a(t)=i\sum_j\,(v^a_j,\partial_{t_k}v^a_j)
  \eqno\enum
$$
has vanishing magnetic flux quantum numbers 
(see sect.~9 for unexplained notations and the relevant lemma).
Once this is achieved the existence of a basis satisfying the 
smoothness condition (8.3) and thus of a smooth measure is guaranteed.

If we define the measure term 
$$
  \widetilde{\L}^a_{\eta}=i\sum_j\,(v^a_j,\delta_{\eta}v^a_j)
  \eqno\enum
$$
as usual, with a tilde to distinguish it from 
the  linear functional $\L_{\eta}$,
the connection may be represented through
$$
  \Bfield_k^a(t)=\widetilde{\L}^a_{\eta},
  \qquad
  \eta_{\mu}(x)=-iU(x,\mu)^{-1}\partial_{t_k} U(x,\mu).
  \eqno\enum
$$
The corresponding expression for the field tensor is
$$
  \Gfield_{kl}(t)=\delta_{\eta}\widetilde{\L}^a_{\zeta}-  
                  \delta_{\zeta}\widetilde{\L}^a_{\eta},
  \eqno\enum
$$
where $\zeta_{\mu}(x)$ is defined in the same way as 
$\eta_{\mu}(x)$ with $\partial_{t_k}$ replaced by 
$\partial_{t_l}$.

We now recall that the measure term satisfies 
the local integrability condition (3.12).
The same is true for $\L_{\eta}$ and we thus conclude that
$$
  \Gfield_{kl}(t)=\delta_{\eta}\L_{\zeta}-  
                  \delta_{\zeta}\L_{\eta}.
  \eqno\enum
$$
In particular,
since $j_{\mu}(x)$ is globally defined and smooth,
the magnetic flux integrals (9.6) are equal to zero
and the U(1) bundle associated with 
the basis $v_j^a(x)$ is hence trivial.

\subsection 10.2 Basis transformation

We may now assume that the basis $v^a_j(x)$ 
satisfies the smoothness condition (8.3). 
The associated current
$\tilde{\jmath}_{\mu}(x)$ is then independent of the patch label $a$
and smoothly dependent on the link variables.
Our aim in the following is to prove that
$$
  \widetilde{\L}_{\eta}=\L_{\eta}-ih^{-1}\delta_{\eta}h,
  \eqno\enum
$$
where $h$ is some globally defined smooth phase factor.
It is then evident that a measure with the required properties 
is obtained by performing a basis transformation
$$
  v^a_j(x)\to \cases{v^a_1(x)h & if $j=1$, \cr
                     \noalign{\vskip1ex}
                     v^a_j(x)  & otherwise. \cr}
  \eqno\enum
$$
This measure is, incidentally, uniquely determined
up to constant phase factor in each topological sector,
because any basis transformation which preserves the measure term
has to have constant determinant.

Recalling lemma 9.3 and our discussion above, it is clear that 
$\L_{\eta}$ and $\widetilde{\L}_{\eta}$ are related
by a basis transformation
if (and only if) the associated Wilson lines (8.9) are the same.
The lemma has been formulated for gauge fields on the $n$-dimensional torus,
but it extends to the field manifold $\gfields[m]$
since the factor $\vfields[m]$ is contractible.

There are two different types of Wilson lines that 
have to be computed. The first of them are associated with the 
gauge loops
$$
  U_t(x,\mu)=\Lambda_t(x)V_{[m]}(x,\mu)\Lambda_t(x+\hat{\mu})^{-1},
  \qquad 
  0\leq t\leq 2\pi,
  \eqno\enum
$$
in field space, where the transformation $\Lambda_t(x)$ is defined by
$$
  \Lambda_t(x)=\exp\left\{it\delta_{\tilde{x}\tilde{y}}\right\},
  \eqno\enum
$$
with $y$ being some fixed lattice point and $\tilde{x}=x\mod L$ as before.
The curve para\-meter $t$ is just one 
of the coordinates $t_k$ on the torus while all other coordinates are set 
zero. As discussed at the end of sect.~9, it is not necessary to 
work out the Wilson lines at other values of the coordinates 
since they are related to each other through eq.~(9.16).

The other non-contractible loops that we need to consider are given by
$$
  U_t(x,\mu)=V_{[m]}(x,\mu)
  \exp\left\{it\delta_{\mu\nu}\delta_{\tilde{x}_{\nu}0}\right\},
  \qquad 
  0\leq t\leq 2\pi,
  \eqno\enum
$$
where $\nu$ is a fixed index.
A somewhat surprising fact is that the Wilson lines around all
these loops can be computed exactly in terms of the anomaly $\anomaly_L(x)$.
This is so for both currents, $j_{\mu}(x)$ and $\tilde{\jmath}_{\mu}(x)$,
and in the following two subsections we shall show that the Wilson lines 
which one obtains are the same, thus completing the proof of theorem 5.1.

\subsection 10.3 Computation of Wilson lines (gauge loops)

In the case of the gauge loop (10.8) we have 
$$
  \eta_{\mu}(x)=-iU_t(x,\mu)^{-1}\partial_tU_t(x,\mu)=
  -\drv{\mu}\delta_{\tilde{x}\tilde{y}}
  \eqno\enum
$$
and it follows from this and property (d) of the current $j_{\mu}(x)$ that
$$
  \L_{\eta}=\anomaly_L(y).
  \eqno\enum
$$
The anomaly is gauge-invariant and hence independent of $t$. 
We thus obtain
$$
  \Wline=\exp\left\{i\kern0.5pt2\pi\anomaly_L(y)_{t=0}\right\}
  \eqno\enum
$$
for the Wilson line associated with $\L_{\eta}$.

To compute the Wilson line 
associated with the measure term $\widetilde{\L}_{\eta}$
we start from lemma 8.1
and note that the projector $P_t$ is given by
$$
  P_t=R[\Lambda_t]P_0R[\Lambda_t]^{-1}.
  \eqno\enum
$$
The lemma then implies
$$
  \widetilde{\Wline}=\lim_{n\to\infty}
  \det\bigl\{1-P_0+(P_0R[\Lambda_{\Delta t}]^{-1}P_0)^n\bigr\},
  \qquad \Delta t=2\pi/n,
  \eqno\enum
$$
and it is immediately clear from this expression that 
$$
  \widetilde{\Wline}=\exp\bigl\{-i\kern0.5pt2\pi\,
  \Tr_L\bigl[\omega TP_0\bigr]\bigr\}, 
  \qquad \omega(x)=\delta_{\tilde{x}\tilde{y}}.
  \eqno\enum
$$
Recalling the definitions
of the projector $P_0$ and the anomaly $\anomaly_L(x)$, 
it is now obvious that the Wilson lines associated with 
$j_{\mu}(x)$ and $\tilde{\jmath}_{\mu}(x)$ are the same.

\subsection 10.4 Computation of Wilson lines (non-gauge loops) 

To calculate the Wilson lines along the loop (10.10) 
we make use of the symmetry transformation
$$
  U(x,\mu)\to U'(x,\mu)=U(-x-\hat{\mu},\mu)^{-1}.
  \eqno\enum
$$
This is a proper rotation of the lattice which maps the field
tensor $F_{\mu\nu}(x)$ to 
$$
  F'_{\mu\nu}(x)=F_{\mu\nu}(-x-\hat{\mu}-\hat{\nu}).
  \eqno\enum
$$
In particular, the flux sectors $\gfields[m]$ are invariant
under this transformation.

From Lemma 7.1 and the definition (7.10) of $V_{[m]}(x,\mu)$ 
we now infer that
$$
  V'_{[m]}(x,\mu)=\Omega_0(x)V_{[m]}(x,\mu)\Omega_0(x+\hat{\mu})^{-1}
  \eqno\enum
$$
for some gauge transformation function $\Omega_0(x)$ satisfying
$\Omega_0(0)=1$.
It follows from this that the fields $U_t(x,\mu)$ along the 
curve transform according to
$$
  \eqalignno{
  U'_t(x,\mu)&=\Omega_t(x)U_{2\pi-t}(x,\mu)\Omega_t(x+\hat{\mu})^{-1},
  &\enum\cr
  \noalign{\vskip2ex}
  \Omega_t(x)&=\Omega_0(x)\exp\left\{it\delta_{\tilde{x}_{\nu}0}\right\}.
  &\enum\cr}
$$
Up to a gauge transformation
the curve is thus mapped onto itself with the reversed 
orientation. 

Taking the gauge invariance and the other 
symmetry properties of the current $j_{\mu}(x)$ into account,
an immediate consequence of these observations is that 
$$
  \left.j_{\mu}(x)\right|_{t\to2\pi-t}=-j_{\mu}(-x-\hat{\mu}).
  \eqno\enum
$$
Along the curve the field variation $\eta_{\mu}(x)$ is given by
$$
  \eta_{\mu}(x)=-iU_t(x,\mu)^{-1}\partial_tU_t(x,\mu)=
  \delta_{\mu\nu}\delta_{\tilde{x}_{\nu}0}
  \eqno\enum
$$
and it is now straightforward to show that 
$$
  \int_0^{2\pi}\rmd t\,\L_{\eta}=
  \int_0^{\pi}\rmd t\,\sum_{x\in\Gamma}
  \delta_{x_{\nu}0}\kern0.5pt\drvstar{\mu}j_{\mu}(x).
  \eqno\enum
$$
Using property (d) of the current, the result
$$
  \Wline=\exp\biggl\{i\int_0^{\pi}\rmd t\,\sum_{x\in\Gamma}
  \delta_{x_{\nu}0}\kern0.5pt\anomaly_L(x)\biggr\}
  \eqno\enum
$$  
is thus obtained.

Our starting point for the computation of the Wilson line
$\widetilde{\Wline}$ associated with the measure term
is again lemma 8.1. The symmetry discussed above implies that 
$$
  P_tQ_t=Q_tP_{2\pi-t},
  \qquad
  Q_tQ_{2\pi-t}=1,
  \eqno\enum
$$
where $Q_t$ is a unitary operator which acts on fermion fields 
according to
$$
  Q_t\psi(x)=R[\Omega_t(-x)]\dirac{5}\psi(-x).
  \eqno\enum
$$
If we set $n=2r$ in eq.~(8.11) it follows from this that 
$$
  \eqalignno{
  \widetilde{\Wline}&=\lim_{n\to\infty}\det\bigl\{
  1-P_0+P_0(Q_{t_1})^{-1}P_{t_1}Q_{t_1}\ldots(Q_{t_r})^{-1}P_{t_r}Q_{t_r}
  \hbox{\kern2em}
  &\cr\noalign{\vskip1ex}
  &\kern5.5em\times P_{t_{r-1}}P_{t_{r-2}}\ldots P_{t_1}P_0\bigr\}.
  &\enum\cr}
$$
We may now insert the representation (8.12) for the projectors and in a few
steps one then ends up with the expression
$$
  \widetilde{\Wline}=\Wline\times
  \det\bigl\{1-P_0+P_0(Q_0)^{-1}P_0\bigr\}
  \det\bigl\{1-P_{\pi}+P_{\pi}Q_{\pi}P_{\pi}\bigr\}.
  \eqno\enum
$$
The operator $Q_0$ maps the space 
of left-handed fields at $t=0$ onto itself and its square is equal
to $1$. Similarly $Q_{\pi}$ operates in the space of left-handed fields
at $t=\pi$ and its square is also equal to $1$. The determinants
in eq.~(10.29) thus contribute to the sign of the Wilson line
and we are left with the problem to prove that the product of 
these sign factors is positive.

First note that the determinants are products of 
$N$ sign factors, one for each fermion flavour.
In all cases where the charge $\rme_{\alpha}$ is even we have
$$
  U_0(x,\mu)^{\rme_{\alpha}}=U_{\pi}(x,\mu)^{\rme_{\alpha}},
  \qquad
  \Omega_0(x)^{\rme_{\alpha}}=\Omega_{\pi}(x)^{\rme_{\alpha}},
  \eqno\enum
$$
and the contribution of these fermions is thus equal to $1$.
If one has a pair of charges $\rme_{\alpha}=-\rme_{\alpha'}$ 
one can use charge conjugation to show that the corresponding factors 
cancel each other and the same is trivially true
for any pair of equal charges.
This proves that $\widetilde{\Wline}=\Wline$ if the charges $\rme_{\alpha}$
satisfy the constraint (5.1).

In the vacuum sector $\gfields[0]$ the situation is simplified 
by the fact that 
the sign factors are the same for any pair of odd charges.
The anomaly cancellation condition (1.1) implies
that the total number of odd charges is even so that 
the Wilson lines coincide in the vacuum sector
independently of whether the constraint (5.1) is satisfied or not.

\section 11. Proof of theorem 5.4

\vskip-4.0ex

\subsection 11.1 Properties of\/ $\jinfty_{\mu}(x)$ in finite volume

For any admissible gauge field $U(x,\mu)$ in finite volume,
the current $\jinfty_{\mu}(x)$ is well-defined and periodic in $x$.
The current thus becomes a local composite field 
on the finite lattice and we now proceed to study the associated linear 
functional
$$
  \Lapprox_{\eta}=\sum_{x\in\Gamma}\eta_{\mu}(x)\jinfty_{\mu}(x).
  \eqno\enum
$$
Note that $\Lapprox_{\eta}$ 
is not quite the same as $\Linfty_{\eta}$, since the  
latter is defined for source fields with compact support while 
we here assume that $\eta_{\mu}(x)$ is a periodic field.

To work out the curvature of $\Lapprox_{\eta}$ it is helpful to 
define the truncated fields
$$
  \eta^n_{\mu}(x)=\cases{\eta_{\mu}(x) & if $x-Ln\in\Gamma$, \cr
                         \noalign{\vskip1ex}
                         0             & otherwise, \cr}
  \eqno\enum
$$
for any integer vector $n$. Translation invariance and periodicity 
then imply
$$
  \delta_{\eta}\Lapprox_{\zeta}-\delta_{\zeta}\Lapprox_{\eta}
  =\sum_{n\in\gz^4}
  \bigl\{\delta_{\eta^n}\Linfty_{\zeta^0}
  -\delta_{\zeta^0}\Linfty_{\eta^n}\bigr\}
  \eqno\enum
$$
and after inserting eq.~(5.9) one obtains
$$
  \delta_{\eta}\Lapprox_{\zeta}-\delta_{\zeta}\Lapprox_{\eta}
  =i\kern0.5pt\Tr\bigl\{\chigamma\hat{P}_{-}\bigl[
  \delta_{\eta}\hat{P}_{-},\delta_{\zeta}\hat{P}_{-}\bigr]\bigr\}.
  \eqno\enum
$$
The projector which appears in this equation is defined by
$$
  \chigamma\psi(x)=\cases{\psi(x) & if $x\in\Gamma$, \cr
                      \noalign{\vskip1ex}
                      0             & otherwise, \cr}
  \eqno\enum
$$
and the trace is taken in infinite volume.
Evidently the right-hand sides of eqs.~(3.12) and (11.4) are different 
and $\jinfty_{\mu}(x)$ itself is, therefore, not an acceptable choice for
the current $j_{\mu}(x)$ in finite volume.

We now introduce the kernel
$$
  P(x,y)=\frac{1}{2}(1-\dirac{5})\delta_{xy}+\frac{1}{2}\dirac{5}D(x,y)
  \eqno\enum
$$
of the projector $\hat{P}_{-}$ in infinite volume and note that 
$$
  \eqalignno{
  &\Tr_L\bigl\{\hat{P}_{-}\bigl[
  \delta_{\eta}\hat{P}_{-},\delta_{\zeta}\hat{P}_{-}\bigr]\bigr\}
  =\sum_{x\in\Gamma}\,\sum_{y,z\in\gz^4}\,\sum_{n\in\gz^4}\,
  \tr\bigl\{P(x,y) &\cr
  \noalign{\vskip2ex}
  &\qquad\times\bigl[\delta_{\eta}P(y,z)\delta_{\zeta}P(z,x+Ln)
  -\delta_{\zeta}P(y,z)\delta_{\eta}P(z,x+Ln)\bigr]\bigr\}.
  &\enum\cr}
$$
The trace in eq.~(11.4) coincides with the $n=0$ term in this sum and 
$$
  \R_{\eta\zeta}=i\kern0.5pt\Tr_L\bigl\{\hat{P}_{-}\bigl[
  \delta_{\eta}\hat{P}_{-},\delta_{\zeta}\hat{P}_{-}\bigr]\bigr\}-
  i\kern0.5pt\Tr\bigl\{\chigamma\hat{P}_{-}\bigl[
  \delta_{\eta}\hat{P}_{-},\delta_{\zeta}\hat{P}_{-}\bigr]\bigr\}
  \eqno\enum
$$
is hence equal to the sum of all the other terms.
Using the locality properties of the Dirac operator,
it is possible to deduce the bounds
\footnote{$\dagger$}{\footnotefont
The supremum norm is given by
$\|\eta\|_{\infty}=\sup_{x,\mu}\left|\eta_{\mu}(x)\right|$
and all other norms are the usual ones in the appropriate spaces of indices
}
$$
  \eqalignno{
  \|P(x,y)\|&\leq \kappa_1
  \left(1+\|x-y\|^{\nu_1}\right)
  \rme^{-\|x-y\|/\varrho},
  &\enum\cr
  \noalign{\vskip2ex}
  \|\delta_{\eta}P(x,y)\|&\leq \kappa_2
  \left(1+\|x-y\|^{\nu_2}\right)
  \rme^{-\|x-y\|/\varrho}\|\eta\|_{\infty},
  &\enum\cr}
$$
where $\varrho$ denotes the localization range of the Dirac operator and 
$\kappa_i$ and $\nu_i\geq0$ are some constants that do not 
dependent on the gauge field.
It follows from this that 
$$
  \left|\R_{\eta\zeta}\right|\leq
  \kappa_3L^{\nu_3}\rme^{-L/\varrho}\|\eta\|_{\infty}\|\zeta\|_{\infty}
  \eqno\enum
$$
and the functional $\Lapprox_{\eta}$ thus satisfies the 
integrability condition in finite volume up to 
exponentially small terms.

In the following we construct a gauge-invariant linear functional
$\S_{\eta}$ which transforms in the same way as $\Lapprox_{\eta}$
under the lattice symmetries and which satisfies
$$
  \delta_{\eta}\S_{\zeta}-\delta_{\zeta}\S_{\eta}=\R_{\eta\zeta},
  \qquad
  \left|\S_{\eta}\right|\leq
  \kappa_4L^{\nu_4}\rme^{-L/\varrho}\|\eta\|_{\infty}.
  \eqno\enum
$$
It is then evident that 
$\L_{\eta}=\Lapprox_{\eta}+\S_{\eta}$
is a solution of the integrability condition (3.12) and that the 
associated current $j_{\mu}(x)$ has all the required properties
apart from the fact that condition (d) of theorem 5.1 is not obviously
fulfilled.

\subsection 11.2 Construction of\/ $\S_{\eta}$ at $\Atrans_{\mu}(x)=0$

The submanifold of fields satisfying $\Atrans_{\mu}(x)=0$ is an 
$n$-dimensional torus which may be
parametrized by coordinates $t_1,\ldots,t_n$
as in sect.~9. If we set 
$$
  \Gfield_{kl}(t)=\R_{\eta\zeta},
  \eqno\enum
$$
where $\eta_{\mu}(x)$ and $\zeta_{\mu}(x)$ are the field
variations associated with $t_k$ and $t_l$,
it is clear from the above
that this tensor is exponentially small at large $L$.
Moreover one knows that $\Gfield_{kl}(t)$ is the field tensor of 
a connection on some U(1) bundle over the torus,
because this is the case for both terms in eq.~(11.8).

It follows from this that the flux integrals (9.6) are 
less than $2\pi$ in magnitude and hence equal to zero
if $L$ exceeds a certain multiple of $\varrho$.
On these lattices $\Gfield_{kl}(t)$ thus satisfies 
all premises of lemma 9.2 and we conclude that there exists 
a linear functional $\S_{\eta}$ which satisfies eq.~(11.12) along the 
submanifold of fields with $\Atrans_{\mu}(x)=0$. Note that 
only the variations $\eta_{\mu}(x)$ are admitted here which 
correspond to variations of the parameters $t_1,\ldots,t_n$. 
These are precisely those for which 
$\eta_{\mu}(x)=\Elong_{\mu}(x)$, where
$$
  \Elong_{\mu}(x)=L^{-4}\sum_{y\in\Gamma}\eta_{\mu}(y)
  +\sum_{y\in\Gamma}\drv{\mu}G_L(x-y)\drvstar{\nu}\eta_{\nu}(y)
  \eqno\enum
$$
denotes the longitudinal part of any given variation 
(the Green function $G_L(z)$ has been introduced in sect.~7).

The  solution $\S_{\eta}$ which one obtains
in this way does not have any special symmetry properties,
but since $\R_{\eta\zeta}$ is gauge-invariant and transforms
in the appropriate way under the lattice symmetries,
we can enforce the proper transformation behaviour
by averaging $\S_{\eta}$ over these symmetries and the gauge group.

\subsection 11.3 Extension of\/ $\S_{\eta}$ to all admissible fields

According to lemma 7.4
any given field $U(x,\mu)$ in the sector $\gfields[m]$ may 
be represented in a one-to-one manner by a field
$V_{[m]}(x,\mu)\ring{U}(x,\mu)$ contained
in the submanifold of fields considered above
and the transverse field $\Atrans_{\mu}(x)$.
The field variations $\eta_{\mu}(x)$ accordingly split into a 
longitudinal variation $\Elong_{\mu}(x)$, defined through eq.~(11.14),
and a transverse variation $\Etrans_{\mu}(x)=\eta_{\mu}(x)-\Elong_{\mu}(x)$.

We now consider the curve
$$
  U_t(x,\mu)=V_{[m]}(x,\mu)\ring{U}(x,\mu)\rme^{it\atrans_{\mu}(x)},
  \qquad 0\leq t\leq1,
  \eqno\enum
$$
and define the functional
$$
  \eqalignno{
  \S_{\eta}&=
  \left.\S_{\elong}\right|_{t=0}+
  &\cr\noalign{\vskip2ex}
  &i\int_0^1\rmd t\,
  \Bigl[\Tr_L\bigl\{\hat{P}_{-}
  \bigl[\partial_t\hat{P}_{-},\delta_{\eta}\hat{P}_{-}\bigr]\bigr\}-
  \Tr\bigl\{\chigamma\hat{P}_{-}
  \bigl[\partial_t\hat{P}_{-},\delta_{\eta}\hat{P}_{-}\bigr]\bigr\}\Bigr].
  &\enum\cr}
$$
The first term in this equation coincides with 
the linear functional 
constructed in the preceding subsection.
As for the second term we remark that
the square bracket is proportional to $\R_{\eta\zeta}$ 
with $\zeta_{\mu}(x)=\Atrans_{\mu}(x)$. From eq.~(7.15)
and the fact that the field tensor is bounded by $\epsilon$
one infers that $\|\Atrans\|_{\infty}\leq\kappa_5L^4$
and $\S_{\eta}$ is thus exponentially small.
Following the lines
in part (c) of the proof of theorem~5.3 given in sect.~6,
it is also easy to check
that $\S_{\eta}$ has the right curvature.

\subsection 11.4 Final steps

At this point we have constructed a current $j_{\mu}(x)$
which satisfies the bound (5.10) and 
which fulfils conditions (a)--(c) of theorem 5.1.
To show that the last condition is also fulfilled, we 
substitute $\eta_{\mu}(x)=-\drv{\mu}\omega(x)$ in eq.~(3.12)
and make use of the gauge transformation properties of the current
and the projector to the left-handed fields.
This leads to the identity
$$
  \delta_{\zeta}\left\{\sum_{x\in\Gamma}\,
  \omega(x)\bigl[\drvstar{\mu}j_{\mu}(x)-\anomaly_L(x)\bigr]\right\}=0
  \eqno\enum
$$
from which one infers that
$\drvstar{\mu}j_{\mu}(x)-\anomaly_L(x)$
only depends on the topological sector but not on the 
particular gauge field that has been chosen.
Because of translation invariance a dependence on $x$ is then 
also excluded.

It follows from this and the 
index theorem [\ref{HasenfratzII},\ref{LuscherI}] that
$$
  L^4\bigl[\drvstar{\mu}j_{\mu}(x)-\anomaly_L(x)\bigr]
  =-\sum_{y\in\Gamma}\anomaly_L(y)
  \eqno\enum
$$
is an integer.
On large lattices this integer has to vanish since
$$
  \left|\anomaly_L(x)-\anomaly(x)\right|
  \leq\kappa_6L^{\nu_6}\rme^{-L/\varrho}
  \eqno\enum
$$
and since the anomaly in infinite volume cancels up to a divergence term.
We thus conclude that the current $j_{\mu}(x)$ satisfies condition (d) 
when $L$ exceeds a certain multiple of $\varrho$.

\section 12. Concluding remarks

Chiral gauge theories with anomaly-free multiplets of 
Weyl fermions are well-defined to all orders of perturbation theory,
but it is not obvious that they can be consistently formulated
at the non-perturbative level. 
The construction presented in this paper provides an 
affirmative answer to this question 
for the case of abelian gauge theories.
Moreover it shows that one can introduce a momentum cutoff in
these theories without breaking the gauge invariance
or giving up the requirement of locality,
which has long been thought to be impossible.

While the general structure of the lattice theories
that we have described is simple, 
the definition of the fermion integration measure turned out 
to be non-trivial because of the gauge anomaly.
We have shown that the measure can be characterized through
a local current satisfying certain conditions 
and then gave a constructive proof that these conditions can be fulfilled.
It is easy to convince oneself,
using similar arguments as in the proof of
theorem 5.1, that the current is uniquely determined up to
terms which amount to adding counterterms to the gauge field action
with the appropriate symmetry and locality properties.
In other words, this is the usual regularization 
ambiguity which one has in any lattice theory.

The relative normalizations and phases of the different topological sectors
[the weight factors $\weight[m]$ in eq.~(2.28)]
however remain undetermined at this point.
This problem is not specific to the lattice approach and 
it would be sufficient to know the normalizations in the 
semi-classical approximation in the continuum theory
to be able to fix these factors. What seems to be lacking at present
is a theoretical principle which restricts the relative weights
of the different sectors.

Evidently one would be interested in extending the present work
to non-abelian gauge theories.
While the discussion in sects.~2--4 carries over with
little change, it is not obvious how precisely the anomaly cancellation
works out, because the general structure of 
the non-abelian anomaly is currently not known on the lattice.
Moreover the topology of the field space is presumably
not as simple as in the abelian case and the absence of 
global topological obstructions may consequently be difficult to prove.
It is conceivable, however, that some of these problems can be bypassed
if one succeeds in deriving a closed expression for the imaginary part
of the effective action along the lines of 
refs.~[\ref{AlvarezEtAl},\ref{EriceLectures}].

\vskip1.0ex
I would like to thank Peter Hasenfratz, Pilar Hern\'andez, Karl Jansen,
Ferenc Niedermayer and Peter Weisz for helpful discussions.
I am also grateful for hospitality at the 
Max-Planck-Institute in Munich and the 
Institute for Theoretical Physics at the University
of Bern, where part of this work has been completed.

\appendix A

All fields considered in this paper live on a 
four-dimensional hyper-cubic euclidean lattice
with lattice spacing $a=1$. Flavour indices $\alpha,\beta,\ldots$
run from $1$ to $N$ and 
Lorentz indices $\mu,\nu,\ldots$ from $0$ to $3$.
Unless stated otherwise the Einstein summation convention is applied 
to the latter.
The symbol $\epsilon_{\mu\nu\rho\sigma}$ stands for the totally anti-symmetric
tensor with $\epsilon_{0123}=1$ and $\delta_{xy}$ is equal
to $1$ if $x=y$ and zero otherwise. 
The conventions for the Dirac matrices are 
$$
  (\dirac{\mu})^{\dagger}=\dirac{\mu},
  \qquad
  \{\dirac{\mu},\dirac{\nu}\}=2\delta_{\mu\nu},
  \qquad
  \dirac{5}=\dirac{0}\dirac{1}\dirac{2}\dirac{3}.
  \eqno\enum
$$
In particular, $\dirac{5}$ is hermitean and $(\dirac{5})^2=1$.

The forward and backward nearest-neighbour difference operators 
$\drv{\mu}$ and $\drvstar{\mu}$ 
act lattice functions $f(x)$ according to 
$$
  \eqalignno{
  \drv{\mu}f(x)&=f(x+\hat{\mu})-f(x),
  &\enum\cr
  \noalign{\vskip2ex}
  \drvstar{\mu}f(x)&=f(x)-f(x-\hat{\mu}),
  &\enum\cr}
$$
where $\hat{\mu}$ denotes the unit vector in direction $\mu$.

\appendix B

In view of the discussion in subsect.~2.3, it suffices to 
consider the lattice Dirac operator in infinite volume.
For all admissible gauge fields we then require that the following
properties hold.

\vskip2ex
\noindent
(a)~{\it Locality and differentiability.}
Ideally one would like the Dirac operator to be strictly local, 
which would imply that the non-zero contributions to the sum (2.15)
come from the points $y$ in a finite neighbourhood of $x$. Moreover
the kernel $D(x,y)$ should be a smooth function of 
the gauge field variables residing there. 

This sort of locality is, however,
incompatible with the Ginsparg-Wilson relation [\ref{Horvath}] and
a more general notion of locality is hence adopted here, where the kernel
is allowed to have exponentially decaying tails at large distances
[\ref{Locality},\ref{Niedermayer}].
More precisely we demand that $D$ is a 
sum of strictly local operators,
$$
  D(x,y)=\sum_{k=1}^{\infty}D_k(x,y),
  \eqno\enum
$$
with localization regions whose diameter $d_k$ grows
at most linearly with $k$.
Moreover these kernels and their derivatives 
$D_k(x,y;z_1,\mu_1;\ldots;z_n,\mu_n)$ with respect to the 
gauge field variables $U(z_1,\mu_1),\ldots,U(z_m,\mu_n)$
are required to satisfy the bounds
$$
  \left\|D_k(x,y;z_1,\mu_1;\ldots;z_n,\mu_n)\right\|
  \leq C_nk^{p_n}\rme^{-\theta k},
  \eqno\enum
$$
where the constants $C_n$, $p_n\geq0$ and $\theta>0$ 
are independent of the gauge field.

The important point to note here is that at large separations 
only the terms with large $k$ contribute. As a consequence we have
$$
  \|D(x,y)\|\leq C\left(1+\|x-y\|^p\right)
  \rme^{-\|x-y\|/\varrho}
  \eqno\enum
$$
for some constants $C$ and $p\geq0$.
The localization range
$$
  \varrho=\sup_{k\geq1}\left\{d_k/(\theta k)\right\}
  \eqno\enum
$$
is a fixed number in lattice units and is thus  
microscopically small compared to the physical distances in the theory.
From the point of view of the continuum limit this kind
of locality is hence as good as strict locality.

\vskip2ex
\noindent
(b)~{\it Gauge covariance and lattice symmetries.}
Under gauge transformations and the lattice symmetries 
(translations, rotations, reflections,
charge conjugation), the Dirac operator and the operators defined by
the kernels $D_k(x,y)$ should transform in the same way as the 
Wilson-Dirac operator $D_{\rm w}$ defined below.

\vskip2ex
\noindent
(c)~{\it Free fermion limit.}
When the gauge field is set to the classical vacuum configuration,
$U(x,\mu)=1$, it follows from (a) and (b) that
$$
  D(x,y)=\int_{-\pi}^{\pi}{\rmd^4p\over(2\pi)^4}\,\rme^{ip(x-y)}
  \widetilde{D}(p),
  \eqno\enum
$$
where $\widetilde{D}(p)$ is an analytic function
in the momenta $p_{\mu}$ with period $2\pi$.
To obtain the correct spectrum of fermions
we require that $\widetilde{D}(p)$ is invertible 
for all non-zero momenta (mod $2\pi$), while for $p\to0$ 
it should be equal to $i\dirac{\mu}p_{\mu}+\rmO(p^2)$.
 
\vskip2ex
\noindent
(d)~{\it Chiral symmetry and hermiticity.}
To preserve chiral symmetry on the lattice, the Dirac operator
should be a solution of the Ginsparg-Wilson relation (1.2).
This identity alone does not imply any hermiticity properties of $D$,
but it is consistent to require that $D^{\dagger}=\dirac{5}D\dirac{5}$.

\vskip2ex
\noindent
(e)~{\it Flavour coherence.}
The last requirement is that 
the Dirac operator should be diagonal
in flavour space with diagonal entries that are obtained from 
the same analytic expression by substituting 
$U(x,\mu)\to U(x,\mu)^{\rme_{\alpha}}$ in the sector
with flavour~$\alpha$. 

\vskip2ex
A relatively simple solution of the Ginsparg-Wilson relation 
has been derived by Neuberger by applying the overlap formalism
to vector-like gauge theories [\ref{NeubergerI}]. 
Explicitly this operator is given by
$$
  D=1-A(A^{\dagger}A)^{-1/2},
  \qquad
  A=1-D_{\rm w},
  \eqno\enum
$$
where $D_{\rm w}$ denotes the standard Wilson-Dirac operator
$$
  D_{\rm w}=\frac{1}{2}\left\{\dirac{\mu}(\nabstar{\mu}+\nab{\mu})
  -\nabstar{\mu}\nab{\mu}\right\}.
  \eqno\enum
$$
Note that one has to insert the representation (2.13)
in the definition
$$
  \eqalignno{
  \nab{\mu}\psi(x)&=
  R[U(x,\mu)]\psi(x+\hat{\mu})-\psi(x),
  &\enum\cr
  \noalign{\vskip2ex}
  \nabstar{\mu}\psi(x)&=
  \psi(x)-R[U(x-\hat{\mu},\mu)]^{-1}
  \psi(x-\hat{\mu}),
  &\enum\cr}
$$
of the gauge covariant forward and backward difference operators.
It is trivial to show that Neuberger's operator satisfies (b)--(e)
and property (a) has recently been established 
for small $\epsilon$ [\ref{Locality}].

\vfill\eject


\beginbibliography


\bibitem{Shamir}
Y. Shamir,
Nucl. Phys. B (Proc. Suppl.) 47 (1996) 212


\bibitem{Rome}
A. Borrelli, L. Maiani, G. C. Rossi, R. Sisto and M. Testa,
Phys. Lett. B221 (1989) 360;
Nucl. Phys. B333 (1990) 335

\bibitem{RomeReview}
M. Testa, The Rome approach to chirality,
Talk given at the APCTP - ICTP Joint International Conference (AIJIC 97) 
on Recent Developments in
Nonperturbative Quantum Field Theory, Seoul 1997,
hep-lat/9707007

\bibitem{BuckowReview}
W. Bock, M. F. L. Golterman and Y. Shamir,
Gauge fixing approach to lattice chiral gauge theories,
talk given at the 31st International Ahrenshoop Symposium on the Theory of
Elementary Particles, Buckow 1997,
hep-lat/9804015


\bibitem{IntA}
M. G\"ockeler, A. Kronfeld, G. Schierholz and U. J. Wiese,
Nucl. Phys. B404 (1993) 839

\bibitem{IntB}
P. Hern\'andez and R. Sundrum,
Nucl. Phys. B455 (1995) 287;
{\it ibid\/} B472 (1996) 334

\bibitem{IntC}
G. 't Hooft,
Phys. Lett. B349 (1995) 491

\bibitem{IntD}
G. T. Bodwin,
Phys. Rev. D54 (1996) 6497


\bibitem{Kaplan}
D. B. Kaplan,
Phys. Lett. B288 (1992) 342;
Nucl. Phys. B (Proc. Suppl.) 30 (1993) 597

\bibitem{Overlap}
R. Narayanan and H. Neuberger,
Phys. Rev. Lett. 71 (1993) 3251;
Nucl. Phys. B412 (1994) 574;
{\it ibid\/} B443 (1995) 305 

\bibitem{OverlapII}
S. Randjbar-Daemi and J. Strathdee,
Phys.Lett. B348 (1995) 543;
Nucl. Phys. B443 (1995) 386;
{\it ibid\/} B466 (1996) 335;
Phys. Lett. B402 (1997) 134


\bibitem{Leutwyler}
H. Leutwyler,
Phys. Lett. B152 (1985) 78

\bibitem{AlvarezEtAl}
L. Alvarez-Gaum\'e, S. Della Pietra and V. Della Pietra,
Phys. Lett. B166 (1986) 177;
Commun. Math. Phys. 109 (1987) 691

\bibitem{EriceLectures}
L. Alvarez-Gaum\'e,
An introduction to anomalies, 
Lectures given
at the International School on Mathematical Physics, Erice 1985, 
{\it in:\/} Fundamental problems of gauge field theory,
eds. G. Velo and A. S. Wightman (Plenum Press, New York, 1986)


\bibitem{HasenfratzI}
P. Hasenfratz,
Nucl. Phys. B (Proc. Suppl.) 63A-C (1998) 53;
Nucl. Phys. B525 (1998) 401 

\bibitem{HasenfratzII}
P. Hasenfratz, V. Laliena and F. Niedermayer,
Phys. Lett. B427 (1998) 125

\bibitem{NeubergerI}
H. Neuberger,
Phys. Lett. B417 (1998) 141;
{\it ibid}\/ B427 (1998) 353

\bibitem{LuscherI}
M. L\"uscher,
Phys. Lett. B428 (1998) 342

\bibitem{GinspargWilson}
P. H. Ginsparg and K. G. Wilson,
Phys. Rev. D25 (1982) 2649


\bibitem{KikukawaYamada}
Y. Kikukawa and A. Yamada,
Weak coupling expansion of massless QCD with 
a Ginsparg-Wilson fermion and axial U(1) anomaly,
hep-lat/9806013


\bibitem{Locality}
P. Hern\'andez, K. Jansen and M. L\"uscher,
Locality properties of Neuberger's lattice Dirac operator,
hep-lat/9808010

\bibitem{Horvath}
I. Horvath, Ginsparg-Wilson relation and ultralocality,
hep-lat/9808002


\bibitem{HasenfratzNiedermayer}
P. Hasenfratz and F. Niedermayer,
private communication

\bibitem{Niedermayer}
F. Niedermayer,
Exact chiral symmetry, topological charge and related topics,
Talk given at the International Symposium on Lattice Field Theory,
Boulder 1998, hep-lat/9810026

\bibitem{OverlapSplit}
R. Narayanan,
Phys. Rev. D58 (1998) 97501


\bibitem{LuscherII}
M. L\"uscher,
Topology and the axial anomaly in abelian lattice gauge theories,
hep-lat/9808021


\bibitem{OverlapAnomaly}
H. Neuberger, 
Geometrical aspects of chiral anomalies in the overlap,
\hfill\break
hep-lat/9802033

\endbibliography

\bye